\documentclass[prd,aps,superscriptaddress,nofootinbib,amsmath,amssymb,showpacs,9pt]{revtex4}
\usepackage{graphicx}
\usepackage{braket}
\usepackage{nicefrac}
\usepackage{float}%
\usepackage{amsmath}
\usepackage{hyperref}
\usepackage{multirow}
 \usepackage{bbold} 

\usepackage[usenames, dvipsnames]{color}

\def\slashchar#1{\setbox0=\hbox{$#1$}
   \dimen0=\wd0 \setbox1=\hbox{/} \dimen1=\wd1
   \ifdim\dimen0>\dimen1 \rlap{\hbox to \dimen0{\hfil/\hfil}} #1
   \else  \rlap{\hbox to \dimen1{\hfil$#1$\hfil}} / \fi}
\def\p{\slashchar{p}}
\def\q{\slashchar{q}}

\graphicspath{{./plots/}}

\begin{document}
\title{Weak production of $\eta$ mesons induced by $\nu_{\mu}(\bar{\nu}_\mu)$ at MicroBooNE energies}

\author{A. \surname{Fatima}}
\affiliation{Department of Physics, Aligarh Muslim University, Aligarh-202 002, India}
\author{M. Sajjad \surname{Athar}}
\email{sajathar@gmail.com}
\affiliation{Department of Physics, Aligarh Muslim University, Aligarh-202 002, India}
\author{S. K. \surname{Singh}}
\affiliation{Department of Physics, Aligarh Muslim University, Aligarh-202 002, India}

\begin{abstract} 
We have studied neutral and charged current (anti)neutrino induced $\eta$ production off the free nucleon target at MicroBooNE energies, in the light of recent results reported by the MicroBooNE collaboration for the total $\eta$ production cross section. 
This study has been made using a theoretical model 
in which the weak hadronic current receives contribution from the nonresonant Born terms as well as from the resonance excitations. 
The Born terms are obtained using the SU(3) symmetric chiral model, used earlier in the study of $K-$meson production. 
The contribution from the resonance terms is considered from the excitation of five nucleon resonances viz. $S_{11}(1535)$, $S_{11}(1650)$, $P_{11}(1710)$, 
$P_{11}(1880)$, and $S_{11}(1895)$.  
To fix the parameters of the vector current interaction, this model is first used to study the electromagnetic production of $\eta$ mesons induced by real and virtual photons, and the theoretical results have been compared with the data from the
MAINZ and JLab experiments. The partially conserved axial-vector current
hypothesis and generalized Goldberger-Treiman relation are used to fix the parameters of the axial-vector
current interaction. The results are presented for the total cross section for the neutral and charged current induced $\eta$ production, ratio of the cross sections for the charged current to neutral current, MicroBooNE flux averaged cross section $\langle \sigma \rangle$, $\left \langle \frac{d\sigma}{dQ^2} \right\rangle$ and $\left\langle \frac{d\sigma}{dp_\eta} \right\rangle$, which may be useful in the future analysis of MicroBooNE as well as other accelerator and atmospheric neutrino experiments being performed in the ${\cal O}$(1)~GeV energy region.
\end{abstract}
\maketitle
\section{Introduction}
The study of the weak production of mesons induced by both the charged and neutral currents in the inelastic sector of the (anti)neutrino-nucleon 
interactions has historically been centered around the weak production of pions, which is dominated by the excitation of $\Delta$ resonance and its 
subsequent decay producing pions~\cite{SajjadAthar:2022pjt, Athar:2020kqn}. 
In recent years, the weak production of single pions induced by the charged and neutral currents in the (anti)neutrino reactions has 
attracted considerable interest as it plays very important role in modeling the weak (anti)neutrino-nucleon cross section in the analysis of neutrino oscillation 
experiments in the sub-GeV and few GeV energy regions. 
However, in the GeV energy region of current neutrino oscillation experiments with accelerator neutrinos like MicroBooNE~\cite{MicroBooNE}, 
SBND~\cite{Machado:2019oxb}, T2K~\cite{T2K}, T2HyperK~\cite{Hyper-Kamiokande:2022smq}, 
and DUNE~\cite{DUNE:2022aul} as well as with the atmospheric neutrinos like HyperK~\cite{Hyper-Kamiokande:2018ofw}, JUNO~\cite{JUNO}, and INO~\cite{INO}, the weak 
production of heavier mesons like $K^{\pm}$, $K^{0}(\bar{K}^{0})$, and $\eta$ could also become relevant and would play significant role in modeling 
the neutrino-nucleon cross sections in the inelastic sector of neutrino reactions~\cite{SajjadAthar:2022pjt}. 
Since these heavy mesons are produced by the weak excitation of higher resonances in the strange and nonstrange sectors and their subsequent decays 
into baryons and mesons, in addition to, the nonresonant direct production of mesons, the study of the weak production of heavy mesons provides useful 
information about the electroweak properties of the higher resonances like $S_{11}(1535)$, $D_{13}(1520)$, $S_{11}(1650)$, $P_{11}(1710)$, $\Lambda(1405)$, $\Sigma(1385)$, etc.

In this context while there have been quite a few studies of the weak single and associated production of $K^{\pm}$ and $K^{0}(\bar{K}^0)$ mesons in recent years~\cite{SajjadAthar:2022pjt, Fatima:2020tyh, RafiAlam:2010kf, RafiAlam:2012bro}, there exists very little work on the weak production of $\eta$ mesons. Theoretically, the early work by Dombey~\cite{Dombey:1968vh}, was followed much later by Alam et al.~\cite{Alam:2013vwa} and Nakamura et al.~\cite{Nakamura:2015rta}. 
We have recently studied, in some detail, the weak production of $\eta$ mesons induced by the charged current in the neutrino and antineutrino reactions off the nucleon in the energy region of $E_{\nu(\bar{\nu})} \le 2$~GeV~\cite{Fatima:2022nfn}.  

Experimentally, the first results on the weak production of $\eta$ mesons induced by neutrinos and antineutrinos were reported by the BEBC collaboration~\cite{BEBCWA59:1989ofp} and later by the ICARUS collaboration~\cite{ICARUS}. 
Recently, the MicroBooNE collaboration~\cite{MicroBooNE:2023dqf} has reported the results for $\eta$ production in neutrino interaction on argon by observing two photons through the $\eta \rightarrow 2\gamma$ decay~($\sim 40\%$ B.R.) in the final state with a cross section $\sigma_{\nu \rightarrow 1\eta + X \rightarrow 2\gamma +0 \pi^0 + X} = 1.27 \pm 0.33 \pm 0.34 \times 10^{-41}$ cm$^2$/nucleon,
implying a total cross section $\sigma_{\nu \rightarrow 1\eta + X} = 3.22 \pm 0.84 \pm 0.86 \times 10^{-41}$ cm$^2$/nucleon. 
Since no charged leptons are observed in the final state, this $\eta$ production cross section includes the weak production of $\eta$ mesons induced by the charged 
as well as the neutral current interactions. Further analyses are being done by the MicroBooNE collaboration to isolate the events with a charged lepton in the final state so that the weak $\eta$ 
production induced by charged and neutral currents could be studied separately~\cite{MicroBooNE:2023dqf}. Moreover, the $\nu_{\mu}$ beam at MicroBooNE has contamination by the other neutrino flavors, i.e., $\nu_{\mu}$ being 93.7\%, with 5.8\% of $\bar{\nu}_{\mu}$, 
0.5\% of $\nu_{e}$, and 0.05\% of $\bar{\nu}_{e}$. 
It is, therefore, important to theoretically estimate the $\nu_{l}(\bar{\nu}_{l})~(l=e,\mu)-$nucleon cross section for $\eta$ production induced by the other neutrino flavors in this energy region.
Keeping this in mind, we have studied the weak production of $\eta$ mesons induced by the charged and the neutral weak currents (anti)neutrino--nucleon reactions for all 
the (anti)neutrino flavors, i.e., $\nu_{\mu}$, $\bar{\nu}_{\mu}$, $\nu_{e}$, and $\bar{\nu}_{e}$.
These studies will also be helpful for the future neutrino oscillation programs like DUNE~\cite{DUNE:2022aul} and SBND~\cite{Machado:2019oxb}, in particular, and the other accelerator and atmospheric neutrino experiments being performed in the few GeV energy region, in general.

In our earlier study~\cite{Fatima:2022nfn}, we presented the results for the charged current $\nu_{\mu}(\bar{\nu}_{\mu})$ induced $\eta$ production off the nucleon for $E_{\nu_{\mu}(\bar{\nu}_{\mu})} \le 2$~GeV, 
by taking into account the contribution of the direct nonresonant production and the resonant production due to the excitation and decay of low lying $S_{11}(1535)$, $S_{11}(1650)$, and $P_{11}(1710)$ resonances. 
In this work, we extend our earlier work on the charged current induced $\eta$ production to higher energies  by considering the contribution from additional resonances {viz.} $P_{11}(1880)$ and $S_{11}(1895)$, and also include the weak $\eta$ production due to the neutral current. 
This model is then applied to understand the experimental results from the MicroBooNE collaboration.
The inclusion of the contribution from the higher resonances is needed because the (anti)neutrino flux at the MicroBooNE has a long tail in energy, and the 
flux decreases by two orders of magnitude only beyond $E_{\nu(\bar{\nu})} \ge 2.5$~GeV. Therefore, the flux averaged cross section gets a non-negligible contribution even for $E_{\nu}=2-3$~GeV. It is important to mention that the MicroBooNE flux peaks around $E_{\nu}=0.5-0.6$~GeV with the average energy of the dominant component ($\nu_{\mu}$) flux at 
$E_{\nu_{\mu}} =823$~MeV, while the threshold for the $\nu_{\mu}$ induced charged~(neutral) current $\eta$ production is 880~MeV~(710~MeV). 

The theoretical calculations have been done using the interaction Lagrangian predicted by the standard model~\cite{Weinberg:1967tq, Salam:1968rm} for 
the charged and the neutral current weak interaction of (anti)neutrinos with nucleons.
The contributions from the direct $\eta$ production due to the nonresonant Born diagrams are calculated using a microscopic model based on the SU(3) chiral Lagrangian assuming $\eta$ belonging to the octet representation of SU(3), thus, neglecting the $\eta-\eta^\prime$ mixing. 
The SU(3) Lagrangian has earlier been used to study the weak production of kaons~\cite{SajjadAthar:2022pjt}. 
The contribution from the resonant diagrams due to the excitation of various resonances $R$ like $S_{11}(1535)$, $S_{11}(1650)$, $P_{11}(1710)$, $P_{11}(1880)$, and $S_{11}(1895)$ and their decays into nucleon and $\eta$ through $R\rightarrow N\eta$ mode are calculated using phenomenological Lagrangians where the $\eta$ particle has been treated as the physical meson. 

In the resonance sector, the various parameters appearing in the vector current sector  are fixed by first applying this model to study the photon and  electron induced eta production from the free nucleon. We have fitted the coupling strength at the strong $R \rightarrow N\eta$ vertex and the electromagnetic coupling strength of the $N-R$ transition using the eta photoproduction data on the total cross section available from the MAMI collaboration~\cite{CrystalBallatMAMI:2010slt, A2:2014pie} for $W\le 2$~GeV. Then the $Q^2$ dependence of the electromagnetic $N-R$ transition form factors has been obtained by fitting the data of the electron induced $\eta$ production off the proton target for the total cross section at different values of $Q^2$~($Q^2 <1.4$~GeV$^{2}$) available from the CLAS collaboration~\cite{Denizli:2007tq}. 
In the axial-vector sector, the axial-vector couplings have been calculated using the partially conserved axial-vector current~(PCAC) hypothesis and the generalized Goldberger-Treiman~(GT) relation with inputs from the experimentally determined strong $R \rightarrow N\pi$ couplings. 
In the neutral current induced $\eta$ production, the isospin structure of the neutral currents predicted by the standard model has been used with the experimental values of the  electromagnetic form factors of the nucleon and $N-R$ transition form factors in the electromagnetic sector to determine the weak vector form factors.

Using this model, we have obtained the results for the total scattering cross section for the charged and neutral current $\nu_{\mu} (\bar{\nu}_{\mu})$ and $\nu_{e}(\bar{\nu}_{e})$ induced scattering off the nucleon target, 
the ratio of the total cross section for the charged current to neutral current reactions, 
and finally the MicroBooNE flux averaged $Q^2$ distribution i.e. $\left\langle \frac{d\sigma}{dQ^2}\right \rangle$ vs. $Q^2$, eta momentum distribution i.e. $\left\langle \frac{d\sigma}{dp_{\eta}}\right\rangle$ vs. $p_{\eta}$, and the flux averaged total scattering cross section $\langle\sigma\rangle$.

In Sec.~\ref{sec:eta:EM}, we present the formalism for the photon and the electron induced eta production. In Sec.~\ref{sec:eta:weak}, the formalism for the charged as well as the neutral current $\nu_{l}(\bar{\nu}_{l})~(l=e,\mu)$ induced eta production has been presented. The results and discussions are presented in
Sec.~\ref{results}, and Sec.~\ref{conclude} concludes our findings.

\section{Electromagnetic production of $\eta$ mesons}\label{sec:eta:EM}

\subsection{$\eta$ production induced by photons}\label{sec:eta:photo}

\begin{figure*} 
\begin{center}
\includegraphics[height=8cm,width=0.95\textwidth]{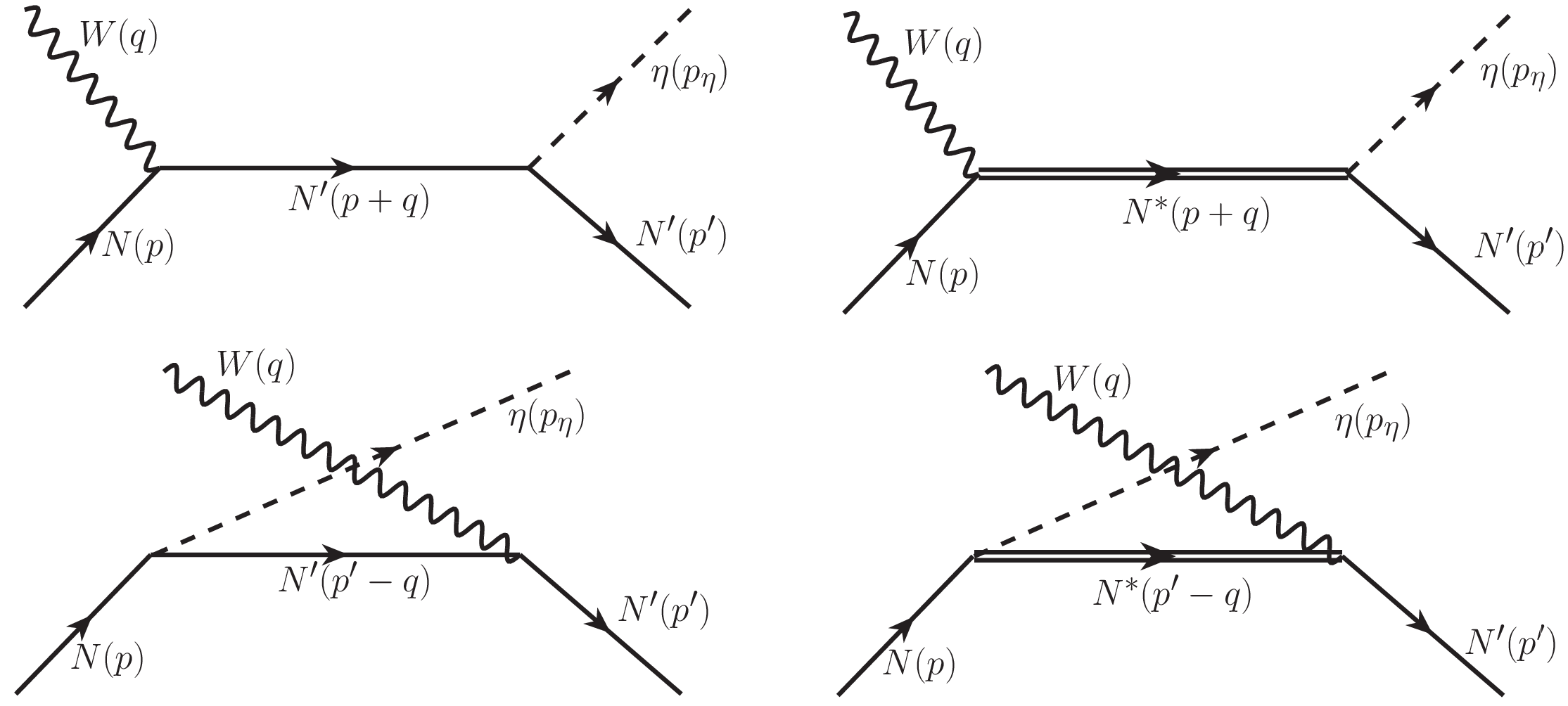}
\caption{Feynman diagrams corresponding to the nonresonant Born terms~(left panel) and resonance excitations~(right panel) for the process $ W (q) + N(p) \longrightarrow \eta(p_{\eta}) + N^\prime(p^{\prime})$. Diagrams shown in the top panel are the nucleon pole diagrams, while the one shown in the bottom panel corresponds to the cross nucleon pole diagrams. In the case of electromagnetic interactions, $W=\gamma,\gamma^{\ast}$ and $N^{\prime} = N = p,n$, while in the case of charged current induced weak interactions, $W=W^{\pm}$ and $N^{\prime}$ and $N$ corresponds to the different nucleons depending upon the charge conservation, and for the neutral current induced reactions, $W=Z$ and $N^{\prime} = N = p,n$. 
The quantities in the parentheses represent the four momenta of the corresponding particles. }\label{Ch12_fg_eta:cc_weak_feynman}
 \end{center}
 \end{figure*}

The differential cross section for the photoproduction of $\eta$ mesons off the free nucleon, {  i.e.},
\begin{equation}\label{eq:eta}
 \gamma(q) + N(p ) \longrightarrow N (p^{\prime}) + \eta(p_{\eta}), \qquad \qquad N=p,n
\end{equation}
is written as~\cite{Fatima:2022nfn}:
\begin{equation}\label{dsig}
\left. \frac{d \sigma}{d \Omega}\right|_{CM} = \frac{1}{64 \pi^{2} s} \frac{|\vec{p}\;^{\prime}|}{|\vec{p}|} 
\overline{\sum_{r}} \sum_{spin} |\mathcal{M}^{r}|^2,
\end{equation}
where the quantities in the parentheses of Eq.~(\ref{eq:eta}) represent the four momenta of the 
corresponding particles. The CM energy $\sqrt{s}$ is expressed as $
 s = W^2 = (q + p)^2 = M^{2} + 2M E_{\gamma} ,$
with $E_{\gamma}$ being the energy of the incoming photon in the laboratory frame. 
$ \overline{\sum} \sum | \mathcal M^{r} |^2$ is the square of the transition matrix element $\mathcal{M}^{r}$, for the photon 
polarization state $r$, averaged and summed over the initial and final spin states, where $\mathcal{M}^{r}$ for reaction~(\ref{eq:eta}) is written in terms of 
the real photon polarization vector $\epsilon_{\mu}^{r}$, as
\begin{equation}
\mathcal{M}^{r} = e \epsilon_{\mu}^{r} (q) J^{\mu} ,
\end{equation}
with $e$ being the electromagnetic coupling constant and $J^{\mu} = \bra{N(p^{\prime}) \eta(p_{\eta})} {J}^{\mu}_{EM} \ket{N(p)}$ being the matrix element
of the electromagnetic current~($J^{\mu}_{EM}$) taken between the hadronic states $\ket{N}$ and $\ket{N\eta}$. The hadronic matrix element  receives contribution from the nonresonant Born terms and the terms corresponding to 
the resonance excitations and their subsequent decay to $N\eta$ mode, which diagrammatically are shown in Fig.~\ref{Ch12_fg_eta:cc_weak_feynman}. 
The hadronic currents for the nonresonant Born terms are 
 obtained using the nonlinear sigma model and the total hadronic current $J^{\mu}$ is  obtained by adding the currents corresponding to the nonresonant and resonance terms coherently. For the detailed description of the formalism, readers are referred to Refs.~\cite{SajjadAthar:2022pjt, Fatima:2022nfn}. 


The expressions of the hadronic currents for $s$- and 
$u$- channels of the $\eta$ photoproduction processes, corresponding to the Feynman diagrams shown in  Fig.~\ref{Ch12_fg_eta:cc_weak_feynman}~(left panel), are obtained as~\cite{SajjadAthar:2022pjt, Fatima:2022nfn}:
\begin{eqnarray}\label{j:s}
J^\mu \arrowvert_{sN} &=&- A_{s}~F_{s}(s) \bar u(p^\prime) \p_{\eta} \gamma_5 \frac{\p + \q + M}
  {s -M^2} \left(\gamma^\mu e_{N} +i \frac{\kappa_{N}}{2 M} \sigma^{\mu \nu} q_\nu \right) u(p), \\
  \label{j:ulam}
J^\mu \arrowvert_{uN} &=&- A_{u} ~F_{u} (u) \bar u(p^\prime) \left(\gamma^\mu e_{N} + i \frac{\kappa_{N}}{2 M} \sigma^{\mu 
\nu} q_\nu \right) \frac{ \p^{\prime} - \q + M}{u - M^2} \p_\eta \gamma_5 u(p),
\end{eqnarray}
where $N$ stands for a proton or a neutron in the initial and final states, $u = 
(p^{\prime} - q)^{2}$, and the strong coupling strengths of $s$ and $u$ channel; $A_{s} = A_{u} = \left(\frac{D - 3F}{2 \sqrt{3} f_{\eta}}\right)$ are obtained using the nonlinear sigma model~\cite{Athar:2020kqn},  assuming the nucleons and the $\eta$ meson belonging, respectively, to the octet baryon and meson representation of the SU(3) representation, thus, neglecting the $\eta - \eta^\prime$ mixing, which is found to be quite small~\cite{DiDonato:2011kr}.
$D$ and $F$ are the axial-vector couplings of the baryon octet and $f_{\eta}=105$~MeV~\cite{Faessler:2008ix} is the $\eta$ decay constant. 

In order to take into account the hadronic structure of the nucleons, the form factors $F_{s} (s)$, and $F_{u} (u)$, are 
introduced at the strong vertex. We use the most general form of the hadronic form factor which is 
taken to be of the dipole form~\cite{Fatima:2020tyh}:
\begin{equation}\label{FF_Born}
F_{x} (x) = \frac{\Lambda_{B}^{4}}{\Lambda_{B}^{4} + (x - M^{2})^{2}}, \qquad \qquad \quad x=s,u
\end{equation}
where $\Lambda_{B}$ is the cut-off parameter taken to be the same for the s- and u-channel nonresonant Born terms, and $x$ represents the Mandelstam variables $s,~u$. The value of $\Lambda_{B}$ 
is fitted to the experimental data for the proton and neutron targets and the best fitted value is 
$\Lambda_{B}=0.75$~GeV and 0.72~GeV, respectively.
In the case of Born terms, the gauge invariance is automatically implemented for the $\eta$ production processes.

In the present work, we have taken into account the resonances, which have mass $M_{R}<2$~GeV and a significant branching ratio to the $N\eta$ decay mode reported in PDG~\cite{ParticleDataGroup:2020ssz}. Specifically, we have considered five spin $\frac{1}{2}$ resonances {  viz.} $S_{11} 
(1535)$, $S_{11} (1650)$, $P_{11} (1710)$, $P_{11}(1880)$, and $S_{11}(1895)$. The general properties of these resonances like mass, decay width, spin, etc. are given in Table~\ref{tab:param-p2}, where we see that $S_{11}(1535)$ resonance dominates the coupling to the $N\eta$ channel.

The most general form of the hadronic currents for the $s-$ and $u-$ channel processes where a resonance state $R_{\frac12}$ 
is produced and decays to a $\eta$ and a nucleon in the final state, are written as~\cite{Athar:2020kqn}:
\begin{eqnarray}
j^\mu\big|_{s}&=& F_{s}^{*} (s) ~\frac{g_{RN\eta}}{f_{\eta}} \bar u({p}\,') 
 \p_{\eta} \gamma_5 \Gamma_{s} \left( \frac{\p+\q+M_{R}}{s-M_{R}^2+ iM_{R} \Gamma_{R}}\right) 
 \Gamma^\mu_{\frac12 
 \pm} u({p}\,), \nonumber\\
 \label{eq:res1/2_had_current}
 j^\mu\big|_{u}&=&  F_{u}^{*} (u) ~\frac{g_{RN\eta}}{f_{\eta}} \bar u({p}\,') 
 \Gamma^\mu_{\frac12 \pm}\left(\frac{\p^{\prime}-\q+M_{R}}{u-M_{R}^2+ iM_{R} \Gamma_{R}}\right) 
 \p_{\eta} \gamma_5 \Gamma_{s}  u({p}\,),
\end{eqnarray}
where $\Gamma_{R}$ and $M_{R}$, respectively, are the decay width and mass of the resonance. $\Gamma_{s} = 1(\gamma_{5})$ stands for the positive~(negative) 
parity resonances, and $g_{RN\eta}$ is the strong coupling strength of the $ R  N\eta$ vertex, which has been determined using the partial decay width of the resonance to $N\eta$ mode where the central values of the full width tabulated in Table~\ref{tab:param-p2} are used in the numerical calculations. The values of the strong coupling constant of the different resonances are also tabulated in Table~\ref{tab:param-p2}. The vertex functions $\Gamma_{\frac{1}{2}^{+}}$ and $\Gamma_{\frac{1}{2}^{-}}$ for the 
positive and negative parity resonances are defined as
\begin{align}\label{eq:vec_half_pos_EM}
  \Gamma^{\mu}_{\frac{1}{2}^\pm} &= {V}^{\mu}_\frac{1}{2}\Gamma_{s},
  \end{align}
where $V^{\mu}_{\frac{1}{2}}$ represents the vector current parameterized in terms of $F_{2}^{R^{+},R^{0}} $, as
 \begin{align}\label{eq:vectorspinhalf1}
  V^{\mu}_{\frac{1}{2}} & =\left[\frac{F_2^{R^{+},R^{0}}}{2 M} 
  i \sigma^{\mu\alpha} q_\alpha \right].
\end{align}
The coupling $F^{R^{+},R^0}_{2}$ is derived from the helicity amplitudes extracted from the real photon scattering 
experiments. The explicit relation between the coupling $F_2^{R^{+},R^0}$ and the helicity amplitude $A_{\frac{1}{2}}^{p,n}$ is given 
by~\cite{Fatima:2022nfn}:
\begin{eqnarray}\label{eq:hel_spin_12}
A_\frac{1}{2}^{p,n}&=& \sqrt{\frac{2 \pi \alpha}{M} \frac{(M_R \mp M)^2}{M_R^2 - M^2}} \left[ \frac{M_R \pm M}{2 M} F_2^{R^{+},R^0} 
\right] ,
\end{eqnarray}
where the upper~(lower) sign stands for the positive~(negative) parity resonance. $R^{+}$ and $R^{0}$ correspond, respectively, to the charged and neutral states of the isospin $\frac{1}{2}$ resonances. 
The value of the helicity amplitude $A_{\frac{1}{2}}^{p,n}$ for the different resonances are  quoted in 
Table~\ref{tab:resonance}.

\begin{table*}
\centering
\begin{tabular*}{180mm}{@{\extracolsep{\fill}}ccccccccc}\hline \hline
&\multicolumn{2}{c}{Resonance $\rightarrow$} & \multirow{2}{*}{$S_{11}(1535)$} & \multirow{2}{*}{$S_{11}(1650)$} & \multirow{2}{*}{$P_{11} (1710)$} & \multirow{2}{*}{$P_{11} (1880)$} &\multirow{2}{*}{$S_{11} (1895)$} &\\
&\multicolumn{2}{c}{Parameters $\downarrow$} &&& &&&\\ \hline
&\multicolumn{2}{c}{$M_{R}$ (GeV)} & $1.510 \pm 0.01$ & $1.655 \pm 0.015$ & $1.700 \pm 0.02$ & $1.860 \pm 0.04$ & $1.910\pm 0.02$ &\\ \hline
&\multicolumn{2}{c}{$\Gamma_{R}$ (GeV)} & $0.130 \pm 0.02$ & $0.135 \pm 0.035$ & $0.120 \pm 0.04$ & $0.230 \pm 0.05$ & $0.110 \pm 0.03$ & \\ \hline
&\multicolumn{2}{c}{$I(J^P)$} &$\frac{1}{2}(\frac{1}{2}^{-})$& $\frac{1}{2}(\frac{1}{2}^{-})$ & $\frac{1}{2}(\frac{1}{2}^{+})$ &$\frac{1}{2}(\frac{1}{2}^{+})$ & $\frac{1}{2}(\frac{1}{2}^{-})$ &\\ \hline 
&\multirow{4}{*}{Branching ratio (in \%)} & $N\pi$ & $32-52$~(43)& $50-70$~(60) & $5-20$~(16) & $3-31$~(34) & $2-18$~(23) &\\ 
&& $N\eta$ & $30-55$~(40) & $15-35$~(25) & $10-50$~(20) & $1-55$~(20) & $15-45$~(30)\\ 
& &$K\Lambda$ &$-$& $5-15$~(10) & $5-25$~(15) & $1-3$~(2) & $3-23$~(13)\\ 
&& $N\pi\pi$ &$4-31$~(17)& $20-58$~(5)& $14-48$~(49)& $>32$~(44) & $17-74$~(34)\\ \hline
&\multicolumn{2}{c}{$|g_{RN\pi}|$} & 0.1019 & 0.0915 & 0.0418 & 0.0466 & 0.0229\\ \hline
&\multicolumn{2}{c}{$|g_{RN\eta}|$} & 0.3696 & 0.1481 & 0.1567 & 0.1369 & 0.0877\\ \hline \hline
\end{tabular*}
\caption{Properties of the spin $\frac{1}{2}$ resonances available in the PDG~\cite{ParticleDataGroup:2020ssz}, with Breit-Wigner mass $M_{R}$, the total decay width $\Gamma_{R}$,
isospin $I$, spin $J$, parity $P$, the branching ratio full range available from PDG~(used in the present calculations) into different meson-baryon channels like $N\pi$, $N\eta$, $K\Lambda$, and $N\pi\pi$, and the strong coupling constant $g_{RN\pi}$ and $g_{RN\eta}$.}
\label{tab:param-p2}
\end{table*}

In analogy with the nonresonant terms,  we have considered the following form factor at the strong 
vertex, in order to take into account the hadronic structure:
\begin{equation}
F^{*}_{x} (x) = \frac{\Lambda_{R}^{4}}{\Lambda_{R}^{4} + (x - M_{R}^{2})^{2}},
\end{equation}
where $\Lambda_{R}$ is the cut-off parameter whose value is fitted to the experimental data. In general, $\Lambda_{R}$ would be different from $\Lambda_{B}$, however, in the case of $\eta$ production by 
photons, it happens that the same value of $\Lambda_{R}$ as that of $\Lambda_{B}$ i.e. $\Lambda_{R} = \Lambda_{B} =
0.75$~GeV for the proton target and $\Lambda_{R} = \Lambda_{B} =
0.72$~GeV for the neutron target gives the best fit to the experimental data. 

To determine the value of the strong $RN\eta$ coupling, we start by writing the most general form of 
$RN\eta$ Lagrangian~\cite{Fatima:2022nfn}:
\begin{align}\label{eq:spin12_lag}
 \mathcal{L}_{R N\eta} &= \frac{g_{ R N\eta} }{f_{\eta}}\bar{\Psi}_{R} \; 
 \Gamma^{\mu}_{s} \;
  \partial_\mu \eta \,\Psi ,
\end{align}
where $\Psi$ is the nucleon field, ${\Psi}_{R}$ 
is the resonance field, and $\eta$ is the eta field. The interaction vertex $\Gamma^{\mu}_{s} = \gamma^\mu \gamma^5$~($\gamma^\mu$) stands for positive~(negative) parity 
resonance states. 

Using the above Lagrangian, one obtains the expression for the decay width in the resonance rest frame as~\cite{SajjadAthar:2022pjt}:
\begin{align}\label{eq:12_width}
 \Gamma_{R \rightarrow N\eta} &= \frac{\mathcal{C}}{4\pi} \left(\frac{g_{RN\eta }}{f_{\eta}}
 \right)^2 \left(M_R \pm M\right)^2 \frac{E_{N} \mp M}{M_R} |\vec{p}^{\,\mathrm{cm}}_{\eta}|,
\end{align}
where the upper~(lower) sign represents the positive~(negative) parity resonance, $\mathcal{C}=1$ for $\eta$ production processes, and $|\vec p^{\,cm}_{\eta}|$ is the 
outgoing eta momentum measured in the resonance rest frame and is given by, 
\begin{equation}\label{eq:pi_mom}
|\vec{p}^{\,\mathrm{cm}}_{\eta}| = \frac{\sqrt{(M_R^2-M_{\eta}^2-M^2)^2 - 4 M_{\eta}^2 M^2}}{2 M_R}  
\end{equation}
and $E_N$, the outgoing nucleon energy is
\begin{equation}\label{eq:elam}
  E_N=\frac{M_R^2+M^2-M_{\eta}^2}{2 M_R}.
\end{equation}

 \begin{figure*}  
\begin{center}
\includegraphics[width=0.85\textwidth,height=8cm]{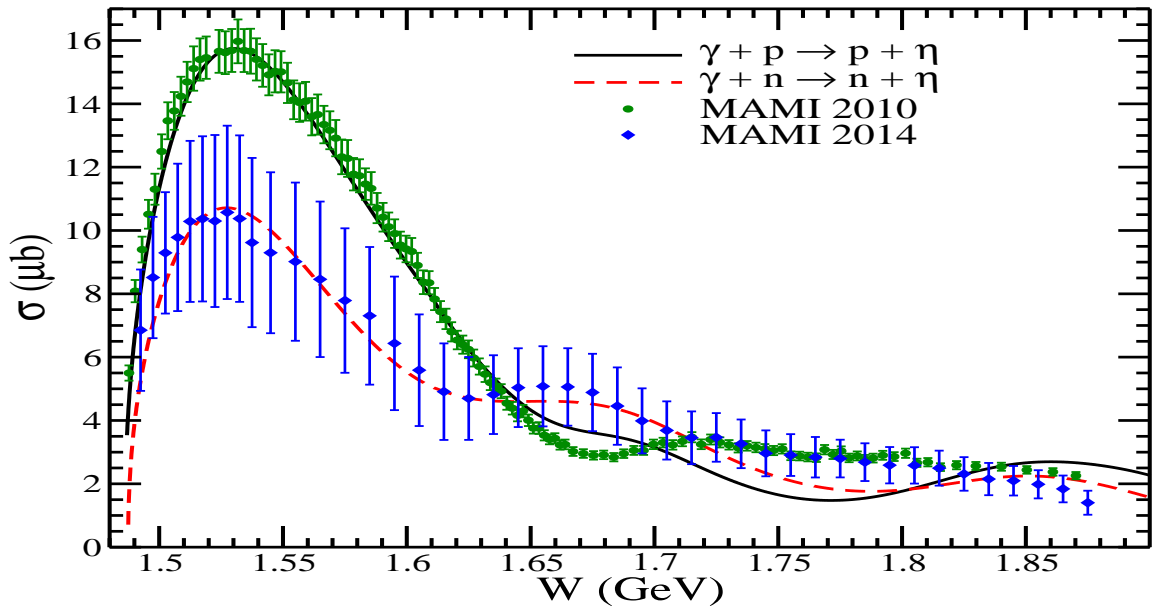}
\caption{Total cross section $\sigma$ vs. $W$ for $\gamma  p \longrightarrow \eta p$~(solid line) 
and $\gamma  n \longrightarrow \eta n$~(dashed line) processes  using the full model. The experimental points for the proton target~(solid circle) are 
obtained from MAMI crystal ball collaboration~\cite{CrystalBallatMAMI:2010slt}, and for the neutron target~(solid diamond) we have used the quasifree neutron data from 
MAMI A2 collaboration~\cite{A2:2014pie}. }
\label{fg_eta:photo_xsec_mami}
\end{center}
\end{figure*}

In Fig.~\ref{fg_eta:photo_xsec_mami}, we have presented the results for the total scattering cross section $\sigma$ as a 
function of $W$ for $\gamma + p \longrightarrow p + \eta$ and $\gamma + n \longrightarrow n + \eta$ 
processes in the region of $W$ from $\eta$ production threshold to $W=1.9$~GeV. We have compared our 
theoretical results with the experimental data obtained by McNicoll et al.~\cite{CrystalBallatMAMI:2010slt}  for the MAMI crystal ball
collaboration on the proton target and the quasifree neutron data from Werthmuller et 
al.~\cite{A2:2014pie} for the MAMI A2 collaboration. 
It may be observed from the figure that in the case of $\eta$ production from the proton and neutron 
targets, our results, with a very few free parameters viz. $\Lambda_B$ and $\Lambda_R$, are in a very good agreement with the 
available experimental data. 

\subsection{Electroproduction of eta meson}\label{sec:eta:elctro}
The electron induced $\eta$ production off the nucleon target is given by the reaction
\begin{equation} \label{eq:elecprod}
  e^- (k) + N(p) \longrightarrow e^-(k^\prime)  + N(p^{\prime}) + \eta (p_{\eta}) \,,
\end{equation}
where the four-momentum for each particle is indicated in the parentheses.
The four-momentum of the virtual photon exchanged in electroproduction is
given by $q = k - k^\prime$. 

The differential scattering cross section for the electroproduction of $\eta$ mesons in the hadronic CM frame is given by~\cite{Fatima:2022nfn}
\begin{equation}\label{dsigma:pion}
\frac{d^5 \sigma}{dE_{l} ~d \Omega_{l} d\Omega_{qp_\eta}} = \frac{1}{32(2\pi)^{5}} \frac{|\vec{k}^{\prime}| |\vec{p}_{\eta}|}{E_{e}M W} \overline{\sum}\sum | \mathcal M |^2,
\end{equation}
where $E_e( E_l)$ is the energy of the incoming~(outgoing) electron; 
$\overline{\sum}\sum | \mathcal M |^2  $ is the square of the transition amplitude averaged~(summed) over the 
spins of the initial~(final) states  with the transition matrix element being written in terms of the leptonic~($l_{\mu}$) and the hadronic~($j^{\mu}$) 
currents as 
\begin{equation}
\label{eq:Gg}
 \mathcal M = \frac{e^2}{{q^2}}\, {l_\mu} j^{\mu}.
\end{equation}
The leptonic current is given as
\begin{equation}\label{eq:lepton_current}
 l_{\mu} = \bar{u} (k^{\prime}) \gamma_{\mu} u(k),
\end{equation}
and $j^{\mu}$ is the sum of the hadronic currents corresponding to the Born terms and resonance excitations, which will be discussed later in this section.

The five-fold differential cross section~(Eq.~\ref{dsigma:pion}) for the electroproduction can also be
expressed as~\cite{Donnachie:1978fm, Amaldi:1979vh, Drechsel:1994zx}:
\begin{equation} \label{eq:5fdxs}
 \frac{d\sigma}{d\Omega_l\, dE_l\, d\Omega_{qp_\eta}}
 = \Gamma\, \frac{d\sigma_\text{v}}{d\Omega_{qp_{\eta}}} \,,
\end{equation}
with the flux of the virtual photon given by
\begin{equation}
 \Gamma = \frac{\alpha}{2 \pi^2}\, \frac{E_l}{E_e}\,
          \frac{K}{Q^2}\, \frac{1}{1-\varepsilon} \,.
\end{equation}
In the above equation, $K =(W^2 - M^2) / 2 M$
denotes the ``photon equivalent energy'', the laboratory energy necessary
for a real photon to excite a hadronic system with CM energy $W$ and $\varepsilon$ is the transverse polarization parameter of the virtual photon, given as
\begin{equation}
 \varepsilon = \left(1 + 2 \frac{|\vec{q}|^2}{\,Q^2}\tan^2\frac{\theta_l}{2}
               \right)^{-1} \,,
\end{equation}
with $Q^2 = - q^2 = -(k-k^\prime)^2$.

The hadronic currents corresponding to the nucleon Born terms exchanged in the $s$- and $u$-channels for the electroproduction of eta mesons, depicted in Fig.~\ref{Ch12_fg_eta:cc_weak_feynman}, are obtained using the nonlinear sigma model and are  written as~\cite{Fatima:2022nfn}: 
\begin{eqnarray}\label{Eq_eta:amp_photo}
J^\mu|_{s(N)} &=&  F_{s}(s)~
\frac{D-3F}{2\sqrt3 f_\eta} \bar u (p^\prime) \slashchar{p_\eta} \gamma^5  
\frac{\slashchar{p}+\slashchar{q}+M}{(p+q)^2-M^2} 
{\cal O}^\mu_N u (p) \nonumber \\ 
J^\mu|_{u(N)} &=&  F_{u}(u)~ \frac{D-3F}{2\sqrt3 f_\eta} 
\bar u (p^\prime) {\cal O}^\mu_N
  \frac{\slashchar{p}-\slashchar{p}_{\eta}+M}{(p - p_\eta)^2-M^2} 
\slashchar{p}_{\eta} \gamma^5 u (p),
\end{eqnarray}
where the 
$\gamma N N$ vertex operator ${\cal O}^\mu_N$ is expressed in terms of the $Q^2$ dependent nucleon form factors as,
\begin{eqnarray}\label{eq:gNN_vertex}
{\cal O}^\mu_N &=& F_1^{N}(Q^2)\gamma^\mu + F_2^{N}(Q^2) i \sigma^{\mu\nu} 
\frac{q_\nu}{2M} .
\end{eqnarray}
The Dirac and Pauli form factors of the nucleon viz. $F_{1}^{p,n}(Q^2)$ and $F_{2}^{p,n} (Q^2)$, respectively, are expressed in terms of the 
Sach's electric~($G_E^{p,n}(Q^2)$) and magnetic~($G_M^{p,n}(Q^2)$) form factors of the nucleons, for which various parameterizations are available in the literature. In the present work we have taken 
 the parameterization of these form factors from Bradford et al.~\cite{Bradford:2006yz} also known as BBBA05 parameterization. For details, see Ref.~\cite{Fatima:2022nfn}.

\begin{table*}
\centering
\begin{tabular*}{160mm}{@{\extracolsep{\fill}}ccc c c c c  c c c}\hline\hline
&Resonance & Helicity amplitude & \multicolumn{3}{c}{ Proton target } & \multicolumn{3}{c}{ Neutron target } &\\ \hline
&&& ${\cal A}_{\alpha} (0)$ & $a_1$ & $b_{1}$ & ${\cal A}_{\alpha} (0)$ & $a_1$ & $b_{1}$ &\\ \hline
&\multirow{2}{*}{$S_{11}(1535)$} & $A_{\frac{1}{2}}$ & 95.0 & 0.85 & 0.85 & $-78.0$ & 1.75 & 1.75& \\ 
&&$ S_{\frac{1}{2}}$ & $-2.0$ & 1.9 & 0.81 & $32.5$ & 0.4 & 1.0& \\ \hline
&\multirow{2}{*}{$S_{11}(1650)$} & $A_{\frac{1}{2}}$ & 33.3 & 0.45 & 0.72 & $26.0$ & 0.1 & 2.5& \\ 
&&$ S_{\frac{1}{2}}$ & $2.5$ & 1.88 & 0.96 & $3.8$ & 0.4 & 0.71& \\ \hline
&\multirow{2}{*}{$P_{11}(1710)$} & $A_{\frac{1}{2}}$ & 55.0 & 1.0 & 1.05 & $-45.0$ & $-0.02$ & 0.95& \\ 
&&$ S_{\frac{1}{2}}$ & $4.4$ & 2.18 & 0.88 & $-31.5$ & 0.35 & 0.85& \\ \hline

&\multirow{2}{*}{$P_{11}(1880)$} & $A_{\frac{1}{2}}$ & $-60.0$ & 0.4 & 1.0 & $-45.0$ & $-0.02$ & 0.95& \\ 
&&$ S_{\frac{1}{2}}$ & $0.4$ & 0.75 & 0.5 & $-31.5$ & 0.35 & 0.85& \\ \hline
&\multirow{2}{*}{$S_{11}(1895)$} & $A_{\frac{1}{2}}$ & $-15.0$ & 1.45 & 0.6 & $26.0$ & $0.1$ & 2.5& \\ 
&&$ S_{\frac{1}{2}}$ & $-3.5$ & 0.88 & 0.6 & $3.8$ & 0.4 & 0.71& \\
\hline \hline
\end{tabular*}
\caption{Parameterization of the helicity amplitude for $S_{11} (1535)$, $S_{11}(1650)$, $P_{11}(1710)$, $P_{11}(1880)$, and $S_{11}(1895)$ resonances on the proton and neutron targets.  ${\cal A}_{\alpha} (0)$ is given in units of $10^{-3}$ GeV$^{-2}$ and the coefficients $a_1$  and $b_1$ in units of GeV$^{-2}$.}
\label{tab:resonance}
\end{table*}

 The general expression of the hadronic current for the resonance excitation in the s- and u- channels, corresponding to the Feynman diagrams shown in Fig.~\ref{Ch12_fg_eta:cc_weak_feynman}~(right panel), is written as,
\begin{eqnarray}\label{jhad:eta_electro}
j^\mu\big|_{s}&=& F_{s}^{*}(s)~ \frac{g_{RN\eta}}{f_{\eta}} \bar u({p}\,') 
 \p_{\eta} \gamma_5 \Gamma_{s} \left( \frac{\p+\q+M_{R}}{s-M_{R}^2+ iM_{R} \Gamma_{R}}\right) 
 \Gamma^\mu_{\frac12 
 \pm} u({p}\,), \nonumber\\
 j^\mu\big|_{u}&=& F_{u}^{*}(u)~ \frac{g_{RN\eta}}{f_{\eta}} \bar u({p}\,') 
 \Gamma^\mu_{\frac12 \pm}\left(\frac{\p^{\prime}-\q+M_{R}}{u-M_{R}^2+ iM_{R} \Gamma_{R}}\right) 
 \p_{\eta} \gamma_5 \Gamma_{s}  u({p}\,).
\end{eqnarray}
The vertex function $\Gamma^\mu_{\frac12 \pm}$ for the positive and negative parity resonances is given in Eq.~(\ref{eq:vec_half_pos_EM}), where the vector current $V_{\frac{1}{2}}^{\mu}$ in the case of electroproduction processes is expressed in terms of the $Q^2$ dependent form factors $F^{R^+,R^0}_{1,2} (Q^2)$ as:
\begin{eqnarray}\label{eq_eta:nstar_em_vertex}
V_{\frac{1}{2}}^{\mu} &=& \frac{F_1^{R}(Q^2)}{(2 M)^2}(\slashchar{q} q^\mu+Q^2\gamma^\mu) 
+ \frac{F_2^{R}(Q^2)}{2 M} i \sigma^{\mu\nu} q_\nu , \qquad R=R^+,R^0  .
\end{eqnarray} 
The electromagnetic transition form factors  for the charged~($F_{1,2}^{R^+}(Q^2)$)  and neutral~($F_{1,2}^{R^0} (Q^2)$) states are then  related to the helicity amplitudes given by the following relations~\cite{SajjadAthar:2022pjt}: 
\begin{eqnarray}\label{eq2}
 A_{\frac{1}{2}}&=&\sqrt{\frac{2\pi\alpha}{K_R}} 
 \Bra{R,J_Z=\frac{1}{2}}\epsilon_\mu^{+} J_i^\mu \Ket{N,J_Z=\frac{-1}{2}}\zeta \nonumber \\ 
 S_{\frac{1}{2}}&=&-\sqrt{\frac{2\pi\alpha}{K_R}}\frac{|\vec q|}{\sqrt{Q^2}} 
\Bra{R,J_Z=\frac{1}{2}}\epsilon_\mu^{0} J_i^\mu \Ket{N,J_Z=\frac{-1}{2}}\zeta
\end{eqnarray}
where in the resonance rest frame, 
\begin{eqnarray}\label{eq1}
K_R&=&\frac{M_R^{2}-M^2}{2M_R}, \quad \qquad |\vec q |^2=\frac{(M_R^{2}-M^{2}-Q^{2})^2}{4M_R^{2}}+Q^2,\nonumber \\
\epsilon^\mu_{\pm}&=&\mp\frac{1}{\sqrt{2}}(0,1,\pm i,0),\qquad \qquad
\epsilon^\mu_{0}=\frac{1}{\sqrt{Q^{2}}}(|\vec q|,1,0 ,q^0).
\end{eqnarray}
The parameter $ \zeta $ is model dependent which is related to the sign of $R \rightarrow N \pi$, 
and for the present calculation is taken as $\zeta =1$. 

\begin{figure*}
\begin{center}
\includegraphics[width=1.\textwidth,height=1.3\textwidth]{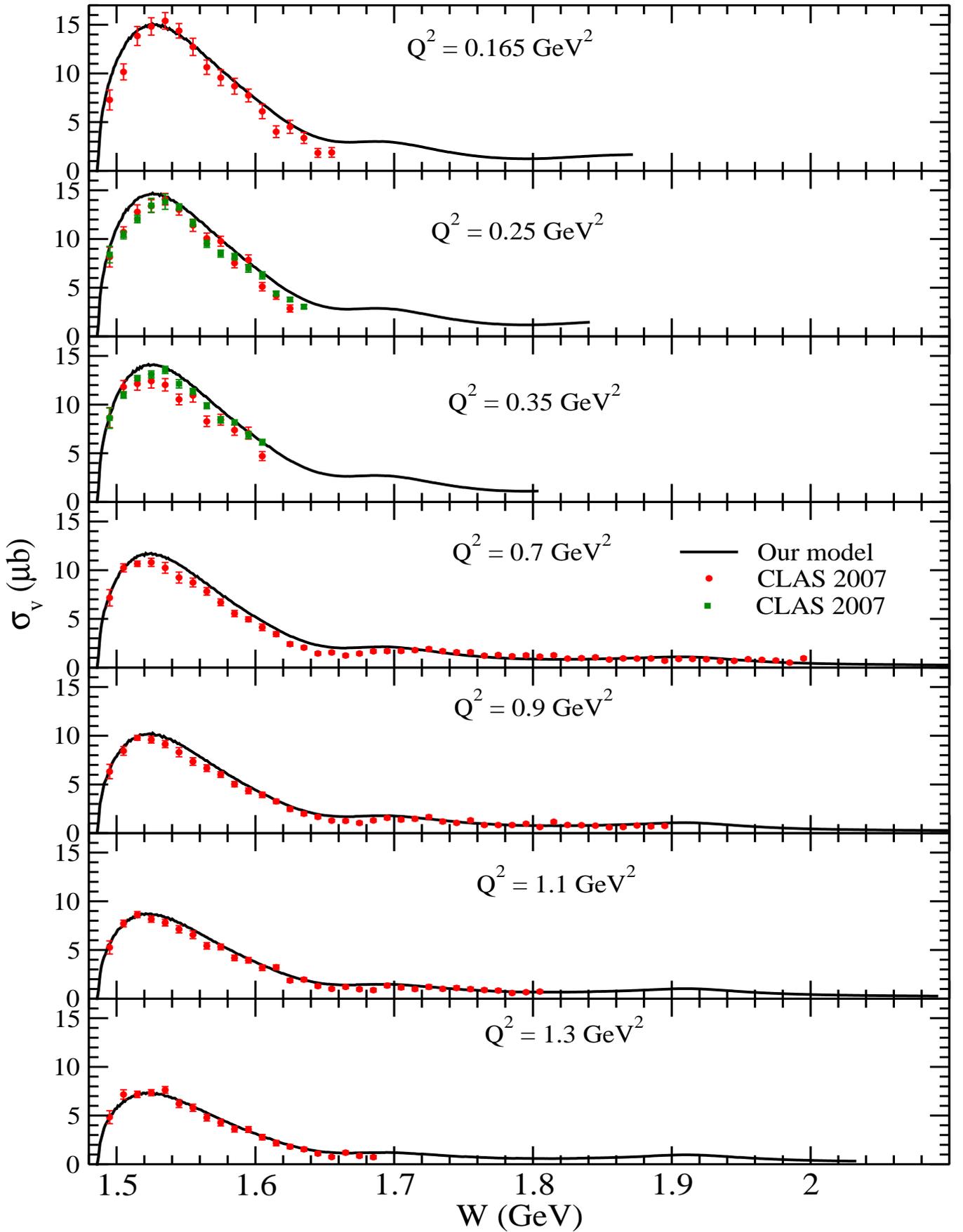}
\caption{Integrated cross section $\sigma_\text{v}$ vs $W$ at different $Q^2$ for $\gamma^\ast  p \rightarrow \eta p$ process. The experimental 
points are the CLAS 2007 data~\cite{Denizli:2007tq}. 
Solid line shows the result of the full model which receives contribution from the nonresonant 
Born terms as well as from the nucleon resonance excitations. 
}
\label{fg_eta:electro_xsec_class}
\end{center}
\end{figure*}

Using Eq.~(\ref{eq1}) in Eq.~(\ref{eq2}), the helicity amplitudes $A_{\frac{1}{2}} (Q^2)$ and $S_{\frac{1}{2}} (Q^2)$ in terms of the electromagnetic
form factors
 $F_1^{R^+,R^0}$ and $F_2^{R^+,R^0}$ are obtained as~\cite{Fatima:2022nfn}:
\begin{eqnarray}\label{eq:hel_em_ff}
A_{\frac{1}{2}}^{p,n} (Q^2)&=& \sqrt{\frac{2 \pi \alpha
}{M}\frac{(M_R \mp M)^2+Q^2}{M_R^2-M^2}} 
\left( \frac{Q^2}{4M^2} F_1^{R^+,R^0}(Q^2) + \frac{M_R \pm M}{2M} F_2^{R^+,R^0}(Q^2) \right) 
\nonumber \\
S_{\frac{1}{2}}^{p,n} (Q^2)&=& \mp \sqrt{\frac{\pi \alpha }{M}\frac{(M_R \pm 
M)^2+Q^2}{M_R^2-M^2}} 
\frac{(M_R \mp M)^2+Q^2}{4 M_R M} \left(  \frac{M_R \pm M}{2M} F_1^{R^+,R^0}(Q^2) - F_2^{R^+,R^0}(Q^2) \right),
\end{eqnarray}
where upper~(lower) sign corresponds to positive~(negative) parity resonances.

The $Q^2$ dependence of the helicity amplitudes~(Eq.~(\ref{eq:hel_em_ff}))  is generally parameterized as~\cite{Tiator:2011pw}:
\begin{equation}\label{eq:ffpar}
{\mathcal A}_{\alpha}(Q^2) = {\mathcal A}_{\alpha}(0) (1+\alpha Q^2)\, e^{-\beta Q^2} ,
\end{equation}
where $ {\mathcal A}_{\alpha}(Q^2)$ are the helicity amplitudes; $A_{\frac12}(Q^2)$ and $S_{\frac12}(Q^2)$ and parameters 
${\mathcal A}_{\alpha}(0)$ are generally determined by a fit to the photoproduction data of the corresponding resonance. 
In the present work, the values of $A_{\frac{1}{2}} (0)$ are taken from  the PDG~\cite{ParticleDataGroup:2020ssz}. 
While the parameters $\alpha$ and $\beta$ are obtained by fitting the electroproduction data on the total cross section at different 
$Q^2$ available from the CLAS experiment~\cite{Denizli:2007tq}, and the values of these parameters for the different nucleon resonances are tabulated in Table~\ref{tab:resonance}.

We obtain the total cross section 
$\sigma_\text{v}$ for $\gamma^\ast  p \rightarrow \eta p$ process by integrating the angular 
distribution~($\frac{d\sigma_\text{v}}{d\Omega_{qp_{\eta}}}$) given in Eq.~(\ref{eq:5fdxs})
over the polar and 
azimuthal angles, which is presented in Fig.~\ref{fg_eta:electro_xsec_class} as a function of CM 
energy $W$ at different values of $Q^2$ ranging from $Q^2= 0.165$~GeV$^{2}$ to 1.3~GeV$^2$. The 
theoretical calculations are presented for the full model, which receives contribution from the nonresonant 
Born terms as well as from the $S_{11}(1535)$, $S_{11} (1650)$, $P_{11}(1710)$, $P_{11}(1880)$ and 
$S_{11}(1895)$ resonance excitations. We have compared our theoretical calculations with the experimental data 
available from the CLAS 
experiment~\cite{Denizli:2007tq} and found a very good agreement between the experimental and 
theoretical results at all values of $Q^2$, including $Q^2>1$~GeV$^{2}$. 

\section{Weak production of $\eta$ mesons}\label{sec:eta:weak}
\subsection{Charged current induced reactions}
The charged current~(CC) (anti)neutrino induced single $\eta$ production off the nucleon 
target~(Fig.~\ref{Ch12_fg_eta:cc_weak_feynman}) are given by the following reactions
\begin{eqnarray}\label{Ch12_eq:eta_weak_process_cc}
\nu_\mu (k)  + n (p) &\longrightarrow& \mu^- (k^\prime) + \eta ( p_\eta) + p (p^\prime),   \\
\label{Ch12_eq:eta_weak_process_cc1}
\bar \nu_\mu (k)  + p (p) &\longrightarrow& \mu^+ (k^\prime) + \eta( p_\eta)  + n(p^\prime) ,
\end{eqnarray}
where the quantities in the parentheses are the four momenta of the particles.

The double differential scattering cross section $\frac{d^2\sigma}{dQ^2 dW}$, for the reactions shown in 
Eqs.~(\ref{Ch12_eq:eta_weak_process_cc}) and (\ref{Ch12_eq:eta_weak_process_cc1}), in the laboratory frame, is expressed as~\cite{Fatima:2022nfn}
\begin{equation}\label{sigma:weak}
 \frac{d^2\sigma}{dQ^2 dW} = \int_{0}^{2\pi} d\phi_{qp_{\eta}} \int_{E_{\eta}^{min}}^{E_{\eta}^{max}} dE_{\eta} \frac{1}{(2\pi)^{4}} \frac{1}{64E_{\nu}^{2}M^2} \frac{W}{|\vec{q}\;|} \overline{\sum} \sum |{\cal M}|^2.
\end{equation}
 The transition matrix element ${\cal M}$, in the case of weak charged current induced process, is given by
\begin{equation}\label{eq:mat:weak}
 {\cal M} = \frac{G_{F}}{\sqrt{2}} \cos\theta_{C}  l_{\mu} J^{\mu},
\end{equation}
with $G_{F}$ being the Fermi coupling constant and $\theta_{C}$ being the Cabibbo mixing angle.
The leptonic current $l_{\mu}$ is given 
\begin{equation}\label{lepton:current}
 l_{\mu} = \bar{u} (k^{\prime}) \gamma_{\mu} (1\mp\gamma_5) u(k)
\end{equation}
where $-(+)$ stands for neutrino~(antineutrino) induced reactions 
and $J^{\mu}= J^{\mu}_{NR} + J^{\mu}_{R}$ is the weak hadronic current, which receives contribution from both the nonresonant Born terms as well as the resonance excitations.

The hadronic currents for the Born diagrams~(s- and u-channels) with nucleon 
poles are given in Eq.~(\ref{Eq_eta:amp_photo}), except for the fact that ${\cal O}_{N}$ is now replaced by ${\cal O}_V$,
where ${\cal O}_{V} = V^{\mu} -A^{\mu}$ is the weak vertex factor. $V^{\mu}$ and $A^{\mu}$ are defined 
in terms of the weak vector and axial-vector form factors as
\begin{align}  \label{eq:vectorspinhalfcurrent}
  V^{\mu}& ={f_{1}^{V}}(Q^2) \gamma^\mu + \frac{f_2^{V}(Q^2)}{2 M} 
  i \sigma^{\mu\nu} q_\nu ,  \\
    \label{eq:axialspinhalfcurrent}
  A^{\mu} &=  \left[{g_1}(Q^2) \gamma^\mu  +  \frac{g_3(Q^2)}{M} q^\mu\right] \gamma_5 ,
\end{align} 
where $f_{1,2}^V(Q^2)$ are, respectively, the isovector vector form factors, and $g_1(Q^2)$ and $g_3(Q^2)$ are 
the axial-vector and pseudoscalar form factors. 
 The two isovector form factors 
$f_{1,2}^V(Q^2)$ are expressed in terms of the Dirac~($F_1^{p,n} (Q^2)$) and Pauli~($F_2^{p,n} (Q^2)$) form factors, discussed in Section~\ref{sec:eta:elctro}, for the 
proton and the neutron,  using the relationships
\begin{equation}\label{Eq_eta:f1v_f2v}
f_{1,2}^V(Q^2)=F_{1,2}^p(Q^2)- F_{1,2}^n(Q^2). 
\end{equation}
These electromagnetic form factors may be rewritten in terms of the electric~($G_{E}^{N} (Q^2)$) and magnetic~($G_{M}^{N} (Q^2)$) Sachs' form factors.
 
The axial-vector form factor $g_1(Q^2)$ is parameterized as
\begin{equation}\label{Eq_eta:fa}
g_1(Q^2)=g_A(0)~\left[1+\frac{Q^2}{M_A^2}\right]^{-2},
\end{equation}
where $g_A(0)=1.267$ is the axial-vector charge and $M_A$ is the axial dipole mass, which in the numerical calculations is taken as the world average value i.e. $M_A = 1.026$~GeV~\cite{Bernard:2001rs}.
On the other hand pseudoscalar form factor $g_3(Q^2)$ is expressed  in terms of $g_1(Q^2)$ 
using the PCAC hypothesis and Goldberger-Treiman relation as
\begin{equation}\label{Eq:fp_nucleon}
g_3(Q^2)=\frac{2M^2g_1(Q^2)}{m_\pi^2+Q^2},
\end{equation}
with $m_\pi$ being the pion mass.

Next, we discuss the positive and negative parity resonance excitation mechanism for the weak interaction induced $\eta$ production. 
The general expression of the hadronic current for the $s-$ and $u-$ channel resonance excitations and their subsequent decay to $N\eta$ mode are given in Eq.~(\ref{jhad:eta_electro}), where the vertex factor $\Gamma_{\frac{1}{2}\pm}^{\mu}$ is now written as
\begin{align}\label{eq:vec_half_pos}
  \Gamma^{\mu}_{\frac{1}{2}^+} &= {V}^{\mu}_\frac{1}{2} - {A}^{\mu}_\frac{1}{2},
  \end{align}
  for the positive parity resonance, and as
\begin{align}\label{eq:vec_half_neg}
  \Gamma^{\mu}_{\frac{1}{2}^-} &= \left({V}^{\mu}_\frac{1}{2} - {A}^{\mu}_\frac{1}{2} \right) \gamma_5 ,
  \end{align}
  for the negative parity resonance. The vector and axial-vector vertex factors for the weak charged current interaction processes are given by
\begin{align}  \label{eq:vectorspinhalfcurrent}
  V^{\mu}_{\frac{1}{2}} & =\frac{{f_{1}^{CC}}(Q^2)}{(2 M)^2}
  \left( Q^2 \gamma^\mu + {q\hspace{-.5em}/} q^\mu \right) + \frac{f_2^{CC}(Q^2)}{2 M} 
  i \sigma^{\mu\alpha} q_\alpha ,  \\
    \label{eq:axialspinhalfcurrent}
  A^{\mu}_\frac{1}{2} &=  \left[{g_1^{CC}}(Q^2) \gamma^\mu  +  \frac{g_3^{CC}(Q^2)}{M} q^\mu\right] \gamma_5 ,
\end{align}
where $f_i^{CC}(Q^2)$~($i=1,2$) are the isovector $N-R$ transition form factors which, in turn, are expressed in terms of the 
charged~($F_{i}^{R+} (Q^2)$) and neutral~($F_{i}^{R0} (Q^2)$) electromagnetic $N-R$ transition form factors as:
\begin{equation}\label{eq:f12vec_res_12}
f_i^{CC}(Q^2) = F_i^{R+}(Q^2) - F_i^{R0}(Q^2), \quad \quad i=1,2
\end{equation}
Further, these form factors are related to the helicity amplitudes as discussed in Section~\ref{sec:eta:elctro}.

The axial-vector current consists of two form factors viz. $g_1^{CC}(Q^2)$ and $g_3^{CC}(Q^2)$, which are determined 
assuming the PCAC hypothesis and pion pole dominance 
of the divergence of the axial-vector current through the generalized GT relation for $N 
- R$ transition~\cite{SajjadAthar:2022pjt}. 

The axial-vector coupling $g_{1}^{CC}$ at $Q^2=0$ is obtained as~\cite{Fatima:2022nfn}
\begin{equation}\label{eq:g1_pos}
g_1^{CC}(0)= 2 g_{RN\pi},
\end{equation}
with $g_{RN\pi}$ being the coupling strength for $R \to N\pi$ decay, 
which has been determined by the partial decay width of the resonance and tabulated in Table~\ref{tab:param-p2}. Since no information about the $Q^2$ dependence of 
the axial-vector form factor is known experimentally, therefore, a dipole form is assumed:
\begin{equation}\label{ga:CC}
 g_1^{CC}(Q^2) = \frac{g_1^{CC}(0)}{\left(1+\frac{Q^2}{M_{A}^2}\right)^2},
\end{equation}
with $M_{A}=1.026$~GeV, and the pseudoscalar form factor $g_3^{CC}(Q^2)$  is given by
\begin{equation}\label{eq:fp_res_spinhalf}
g_{3}^{CC}(Q^2) = \frac{(MM_{R}\pm M^{2})}{m_{\pi}^{2}+Q^{2}} g_1^{CC}(Q^2) ,
\end{equation}
where $+(-)$ sign is for positive~(negative) parity resonances. However, the contribution of $g_{3}^{CC} (Q^2)$ being directly proportional to the lepton mass squared is almost negligible.

\subsection{Neutral current induced reactions}
The neutral current~(NC) (anti)neutrino induced single $\eta$ production off the nucleon 
target~(Fig.~\ref{Ch12_fg_eta:cc_weak_feynman}) are given by the 
following reactions
\begin{eqnarray}\label{Ch12_eq:eta_weak_process_nc}
\nu_l (k)  + N (p) &\longrightarrow& \nu_l (k^\prime) + \eta ( p_\eta) + N (p^\prime),   \\
\label{Ch12_eq:eta_weak_process_nc1}
\bar \nu_l (k)  + N (p) &\longrightarrow& \bar{\nu}_l (k^\prime) + \eta( p_\eta)  + N(p^\prime) , \qquad \qquad N=n,p.
\end{eqnarray}

The expression for the double differential scattering cross section $\frac{d^2\sigma}{dQ^2 dW}$ is
given in Eq.~(\ref{sigma:weak}), where the transition matrix element ${\cal M}$, in the case of 
neutral current  induced process, is given by
\begin{equation}\label{eq:mat:weak}
 {\cal M} = \frac{G_{F}}{\sqrt{2}}   l_{\mu} J^{\mu},
\end{equation}
with the leptonic current being the same as defined in Eq.~(\ref{lepton:current}). The structure of 
the total hadronic current $J^{\mu}$ remains the same as in charged current reactions, i.e., 
$J^{\mu} = {J^{\mu}_{NR}}^{NC} + {J^{\mu}_{R}}^{NC}$, however, the individual hadronic currents for the 
nonresonant Born terms and the resonance excitations are now expressed in terms of the neutral current 
form factors, which are discussed briefly in this section. For details, the readers are referred to 
Ref.~\cite{SajjadAthar:2022pjt}.

%
The hadronic currents for the Born diagrams~(s- and u-channels) with nucleon 
poles are given in Eq.~(\ref{Eq_eta:amp_photo}), however, in the case of NC reactions ${\cal O}_{N}$ is replaced by ${\cal O}_V^{NC}$, the weak neutral current vertex, 
where ${\cal O}_{V}^{NC} = {V^{\mu}}^{NC} - {A^{\mu}}^{NC}$ with ${V^{\mu}}^{NC}$ and ${A^{\mu}}^{NC}$ defined 
in terms of the neutral current form factors as~\cite{SajjadAthar:2022pjt, Athar:2020kqn}:
\begin{align}  \label{eq:vectorNC}
  {V^{\mu}}^{NC}& ={\tilde{f}_{1}}(Q^2) \gamma^\mu + \frac{\tilde{f}_2(Q^2)}{2 M} 
  i \sigma^{\mu\nu} q_\nu ,  \\
    \label{eq:axialNC}
  {A^{\mu}}^{NC} &=  \left[{\tilde{g}_1}(Q^2) \gamma^\mu  +  \frac{\tilde{g}_3(Q^2)}{M} q^\mu\right] \gamma_5 ,
\end{align} 
where $\tilde{f}_{1,2}(Q^2)$ are the neutral current vector form factors and are expressed in terms of both the isovector and isoscalar components, and $\tilde{g}_1(Q^2)$ and $\tilde{g}_3(Q^2)$ are 
the axial-vector and pseudoscalar form factors. 

The vector form factors 
$\tilde{f}_{1,2}(Q^2)$ are expressed in terms of the Dirac~($F_1^{p,n} (Q^2)$) and Pauli~($F_2^{p,n} (Q^2)$) form factors of the nucleon, discussed in Section~\ref{sec:eta:elctro},  using the relationships,
\begin{eqnarray}\label{Eq_eta:f1v_f2v:NC} 
\tilde{f}_i^p (Q^2) &=& \left(\frac{1}{2} -2 \sin^{2}\theta_W \right) F_i^{p} (Q^2) - \frac{1}{2} F_i^{n} (Q^2), \\
  \tilde{f}_i^n (Q^2)&=&\left(\frac{1}{2} -2 \sin^{2}\theta_W \right) F_i^{n} (Q^2) - \frac{1}{2} F_i^{p} (Q^2) , \qquad \qquad i=1,2
\end{eqnarray}
where $\theta_{W}$ is the Weinberg angle.
 
The axial-vector form factor, $\tilde{g}_1(Q^2)$ is expressed as
\begin{equation}\label{Eq_eta:fa:NC}
\tilde{g}_1(Q^2)=\pm \frac{1}{2} g_1 (Q^2),
\end{equation}
where $+~(-)$ stands for proton~(neutron) target and  $g_1(Q^2)$ is defined in Eq.~(\ref{Eq_eta:fa}). The contribution of the pseudoscalar form factor to the transition matrix element is proportional to the lepton mass squared, and therefore does not contribute in the case of NC reactions.

Next, we discuss the resonance excitation mechanism for the neutral current induced $\eta$ production. 
The general expression of the hadronic current for the $s-$ and $u-$ channel resonance excitations and their subsequent decay to $N\eta$ mode are given in Eq.~(\ref{jhad:eta_electro}), with the  vertex factor $\Gamma_{\frac{1}{2}\pm}^{\mu}$  defined in Eqs.~(\ref{eq:vec_half_pos}) and (\ref{eq:vec_half_neg}) for the positive and negative parity resonances, respectively, with the modifications $V^{\mu}_{\frac{1}{2}} \rightarrow {V^{\mu}_{\frac{1}{2}}}^{NC}$ and $A^{\mu}_{\frac{1}{2}} \rightarrow {A^{\mu}_{\frac{1}{2}}}^{NC}$ in the case of NC induced reactions. 
  
  The vector and axial-vector vertex factors for the NC induced processes are given by
\begin{align}  \label{eq:vectorspinhalfcurrent}
  {V^{\mu}_{\frac{1}{2}}}^{NC} & =\frac{{f_{1}^{NC}}(Q^2)}{(2 M)^2}
  \left( Q^2 \gamma^\mu + {q\hspace{-.5em}/} q^\mu \right) + \frac{f_2^{NC}(Q^2)}{2 M} 
  i \sigma^{\mu\alpha} q_\alpha ,  \\
    \label{eq:axialspinhalfcurrent}
  {A^{\mu}_\frac{1}{2}}^{NC} &=  \left[{g_1^{NC}}(Q^2) \gamma^\mu  +  \frac{g_3^{NC}(Q^2)}{M} q^\mu\right] \gamma_5 ,
\end{align}
where $f_i^{NC}(Q^2)$~($i=1,2$) are the neutral current $N-R$ transition form factors which, in analogy with the nucleon form factors, are expressed in terms of the 
charged~($F_{i}^{R+} (Q^2)$) and neutral~($F_{i}^{R0} (Q^2)$) electromagnetic $N-R$ transition form factors as:
\begin{eqnarray}\label{eq:f12vec_res_12}
{f_i}^{NC} (Q^2) &=& \left(\frac{1}{2} -2 \sin^{2}\theta_W \right) F_i^{R+} (Q^2) - \frac{1}{2} F_i^{R0} (Q^2), \qquad \qquad \text{for proton target} \\
  {f}_i^{NC} (Q^2)&=&\left(\frac{1}{2} -2 \sin^{2}\theta_W \right) F_i^{R0} (Q^2) - \frac{1}{2} F_i^{R+} (Q^2) , \qquad \qquad \text{for neutron target}.
\end{eqnarray} 

The axial-vector neutral current form factor $g_1^{NC} (Q^2)$ is expressed in terms of $ g_1^{CC}(Q^2)$ as
\begin{equation}
{g}_1^{NC}(Q^2)=\pm \frac{1}{2} g_1^{CC} (Q^2),
\end{equation}
where $+~(-)$ stands for proton~(neutron) target,  $g_1^{CC}(Q^2)$ is defined in Eq.~(\ref{ga:CC}).

\section{Results and Discussion}\label{results}
\subsection{Total and differential scattering cross sections}
\begin{figure*}
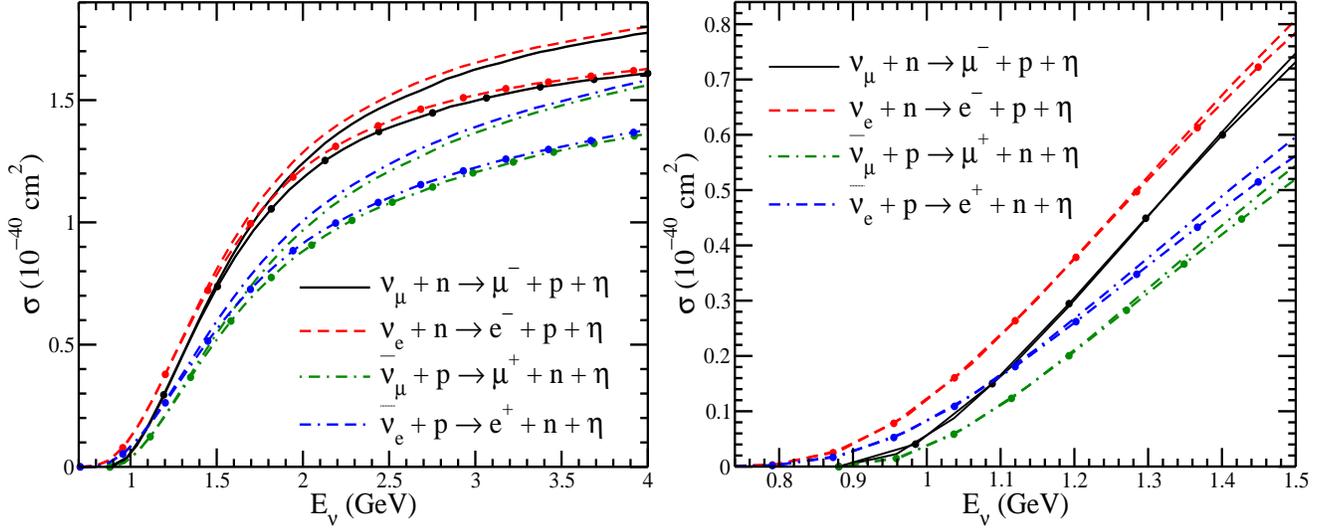

\begin{center}
\includegraphics[width=0.48\textwidth,height=7cm]{total_sigma_e_mu_CC.eps}
\includegraphics[width=0.48\textwidth,height=7cm]{total_sigma_e_mu_CC_15GeV.eps}
\caption{(Left panel)~Total scattering cross section~$\sigma$ for $\nu_{\mu}$~(solid line), $\bar{\nu}_{\mu}$~(dash-dotted line), $\nu_{e}$~(dashed line), and $\bar{\nu}_{e}$~(double-dash-dotted line) CC induced $\eta$ production off the nucleon target as a function of (anti)neutrino energy~($E_{\nu}$), using the full model that receives the contributions from the nonresonant Born terms as well as from the resonance diagrams including $S_{11}(1535)$, $S_{11}(1650)$, $P_{11}(1710)$, $P_{11}(1880)$ and $S_{11}(1895)$. The lines with solid circles show the contribution only from $S_{11}(1535)$ resonance. (Right panel)~Same results but for $E_{\nu}$ from threshold to 1.5~GeV.}\label{Ch12_fg_eta:cc_xsec_weak}
\end{center}
\end{figure*}

\begin{figure*}
\begin{center}
\includegraphics[width=0.48\textwidth,height=7cm]{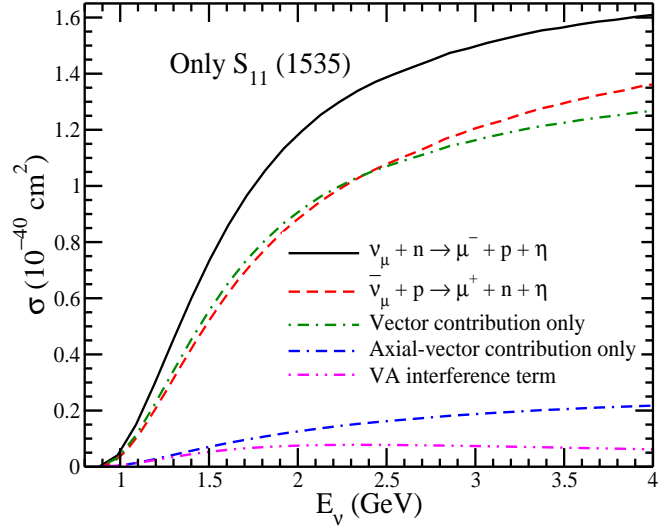}
\caption{Total scattering cross section for the CC induced $\eta$  production i.e. $\nu_{\mu} + n 
\longrightarrow \mu^{-} + \eta + p$~(solid line) and $\bar{\nu}_{\mu} + p \longrightarrow \mu^{+} + \eta + n$~(dashed line) using only the contribution from $S_{11}(1535)$ resonance. Dashed-dotted, double-dashed-dotted, and double-dotted-dashed lines, respectively, show the results for only vector contribution, only axial-vector contribution, and vector axial-vector interference of the weak hadronic current.}\label{Ch12_fg_eta:1535}
\end{center}
\end{figure*}

\begin{figure*}
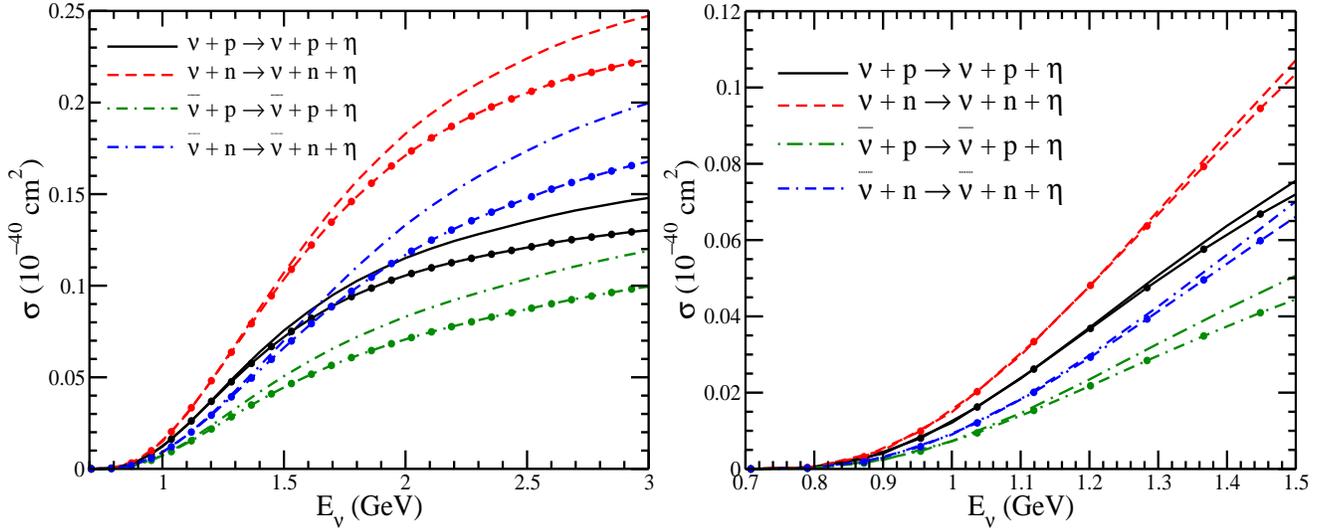

\begin{center}
\includegraphics[width=0.48\textwidth,height=7cm]{total_sigma_NC.eps}
\includegraphics[width=0.48\textwidth,height=7cm]{total_sigma_NC_15GeV.eps}
\caption{(Left panel)~Total scattering cross section for $\nu+p \rightarrow \nu + p + \eta$~(solid line), $\nu + n \rightarrow \nu + n + \eta$~(dashed line), $\bar{\nu} + p \rightarrow \bar{\nu} + p + \eta$~(dash-dotted line), and $\bar{\nu} + n \rightarrow \bar{\nu} + n + \eta$~(double-dash-dotted line) reactions as a function of (anti)neutrino energy. Lines~(lines with solid circles) show the contribution from the full model~($S_{11}(1535)$ resonance only). (Right panel)~Same results but for $E_{\nu}$ from threshold to 1.5~GeV.}\label{Ch12_fg_eta:cc_xsec_NC}
\end{center}
\end{figure*}

\begin{figure*}
\begin{center}
\includegraphics[width=10cm,height=7cm]{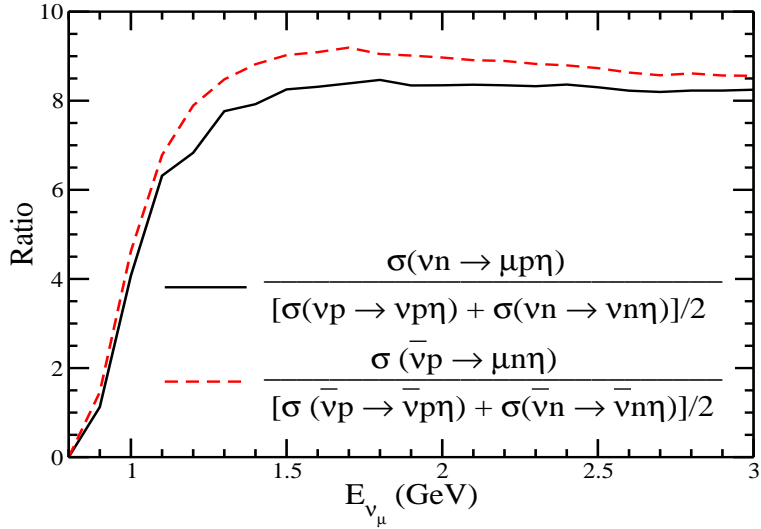}
\caption{Ratio of total scattering cross section for the (anti)neutrino CC induced $\eta$  production to (anti)neutrino NC $\eta$ production off the nucleon target as a function of neutrino energy. Solid line shows the results for the ratio $=\frac{\sigma(\nu_{\mu} + n \rightarrow \mu^{-} + p +\eta)}{\left[\sigma(\nu + p \rightarrow \nu + p + \eta) + \sigma(\nu +  n \rightarrow \nu + n + \eta)\right]/2}$ for the neutrino induced process, and the dashed line shows the results for the ratio $=\frac{\sigma(\bar{\nu}_{\mu} + p \rightarrow \mu^{+} + n +\eta)}{\left[\sigma(\bar{\nu} + p \rightarrow \bar{\nu} + p + \eta) + \sigma(\bar{\nu} +  n \rightarrow \bar{\nu} + n + \eta)\right]/2}$ for the antineutrino induced process.}\label{Ch12_fg_eta:cc_xsec_ratio}
\end{center}
\end{figure*}

 \begin{figure*}
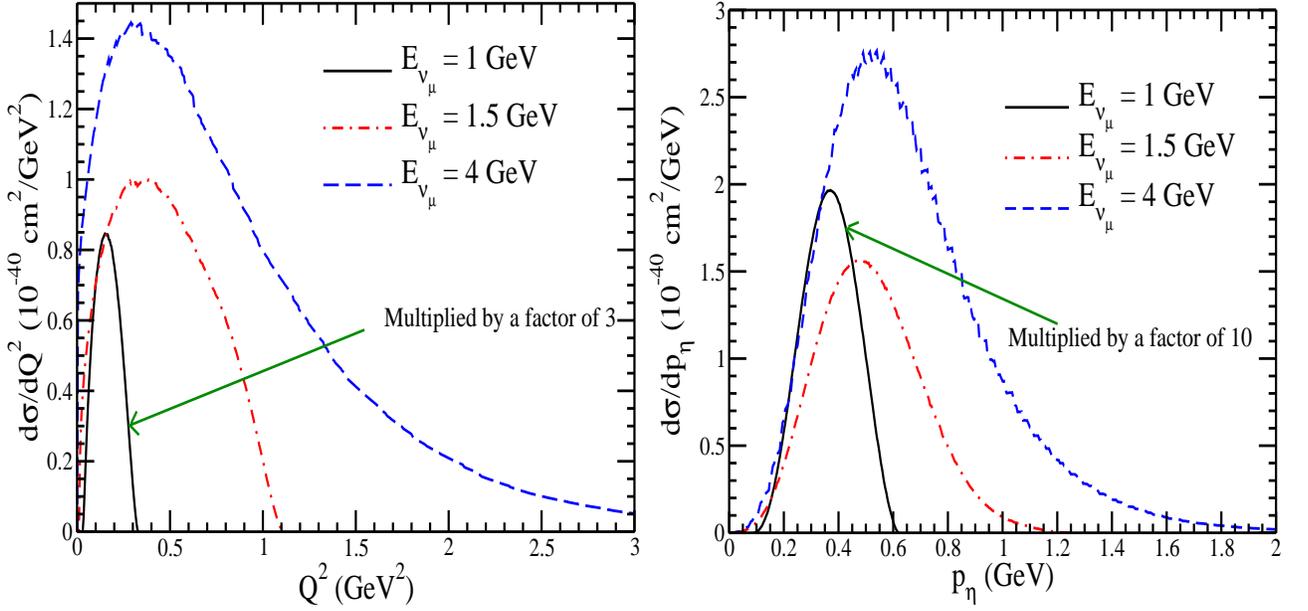

\begin{center}
\includegraphics[width=0.47\textwidth,height=.45\textwidth]{dsigma_dq2_neutrino_different_energy.eps}
\includegraphics[width=0.47\textwidth,height=.45\textwidth]{dsigma_dpeta_neutrino_different_energy.eps}
\caption{$Q^2$ distribution~(left panel) and $\eta$-momentum distribution~(right panel) for the charged current induced $\nu_{\mu} + n \longrightarrow \mu^{-} + p + \eta$ process
at $E_{\nu_\mu({\bar\nu}_\mu)}=1$~GeV~(solid line), 1.5~GeV~(dashed-dotted line) and 4~GeV~(dashed line) using the full model calculation. }
\label{fg_eta:cc_q2_weak}
\end{center}
\end{figure*}

In Fig.~\ref{Ch12_fg_eta:cc_xsec_weak}, we present the results for the total cross section $\sigma$ vs. $E_{\nu_{l}(\bar{\nu}_{l})}$~($l=e,\mu$) for the neutrino and the antineutrino charged current induced $\eta$ production processes. These results are presented both for electron and muon type (anti)neutrinos, by taking into account the contribution from $S_{11}(1535)$ resonance only, and the full model, which includes contribution from the nonresonant Born terms as well as from the resonance excitations.  In the present work, we have considered five spin half resonances viz. $S_{11}(1535)$, $S_{11}(1650)$, $P_{11}(1710)$, $P_{11}(1880)$, and $S_{11}(1895)$. 
It may be observed from the figure that there is a dominance of $S_{11}(1535)$ resonance, which is more pronounced in the case of neutrinos than antineutrinos. 
For example, at $E_{\nu}=1.5$ GeV, the contribution of $S_{11}(1535)$ is 98\% (96\%), which becomes 92\%(89\%) at $E_{\nu}=3$~GeV for neutrino~(antineutrino) induced processes. 
The total contribution of the nonresonant terms is less than 2\% in the energy range $E_{\nu_{\mu}}=1-3$~GeV for the (anti)neutrino induced $\eta$ production processes. 
In view of the small contribution of the nonresonant terms, the assumption of neglecting $\eta-\eta^\prime$ mixing in the evaluation of the nonresonant terms is therefore justified. 
In view of the accelerator experiments like MicroBooNE, T2K, SBND, etc., and  the atmospheric experiments for the sub-GeV energy region, where there is considerable flux of (anti)neutrinos at lower energies~($E_{\nu} \le 1.5$~GeV), we have explicitly shown the dominance of $S_{11}(1535)$ resonance, in the right panel of  Fig.~\ref{Ch12_fg_eta:cc_xsec_weak}, by presenting the results of $\sigma$ as a function of (anti)neutrino energy from threshold up to $E_{\nu}=1.5$ GeV. 
The results obtained in our model are in agreement with the results reported by Nakamura et al.~\cite{Nakamura:2015rta}, in the case of $\nu_{\mu} + n \rightarrow \mu^{-} + p + \eta$ reaction,  using the DCC model and also with our earlier work~(see Fig.~11 of Ref.~\cite{Fatima:2022nfn}).


Since the $\eta$ production cross sections are dominated by $S_{11}(1535)$ resonance, therefore, we have also considered individually the contribution from the vector and axial-vector components of the weak hadronic current due to the $N-S_{11}(1535)$ transition. These results are shown in Fig.~\ref{Ch12_fg_eta:1535}  for $\nu_{\mu}$ and $\bar{\nu}_{\mu}$ induced processes. It may be observed from the figure that the contribution of the vector part of the hadronic current dominates, for example, it has 76\% contribution at 1.5~GeV, which becomes 78\% at 3~GeV. This dominance of the vector contribution is also reported by the very old calculation of Dombey~\cite{Dombey:1968vh}, who finds the ratio of vector to axial-vector contribution to be 2.7:1 at very higher energy, which may be compared with our result of 6.25:1 at $E_{\nu_{\mu}}=4$~GeV.

Since we have fixed the parameters of the vector part of the weak hadronic current by fitting the photo and electroproduction data, therefore, any uncertainty in the cross section for the 
(anti)neutrino induced processes arises mainly due to the uncertainty in the axial-vector part of the weak hadronic current. 
Moreover, the dominant contribution is from the vector current, therefore, the theoretical uncertainty in the total cross section due to the uncertainty in the axial-vector contribution is quite small.
Quantitatively, to understand this uncertainty, we have varied the strong coupling~($g_{RN\pi}$), determined by the partial decay width of $R \rightarrow N\pi$ mode,  maximally allowed by the PDG and found that a 15\% variation in the strong coupling strength results in a  
 change of 3--5\% in the neutrino induced cross section, which is found to be even smaller in the antineutrino induced charged current reactions.
The other uncertainty is due to the axial dipole mass $M_{A}$, the value of which is taken to be equal to $M_{A}=1.026$~GeV. A change of 10\% in $M_{A}$ results a change of 4--6\% in the cross section in the energy range of 1.5 GeV to 3 GeV.

%
 In Fig.~\ref{Ch12_fg_eta:cc_xsec_NC}, we have presented the results for  $\sigma$ vs. $E_{\nu(\bar{\nu})}$ for the neutral current induced (anti)neutrino scattering off proton and neutron targets. These results are presented by taking the contribution from the full model, and from $S_{11}(1535)$ resonance only.  It may be noticed that the total cross section from neutron target is more than the proton target both in the neutrino and antineutrino induced reactions. We find that for neutrino induced reaction from the neutron, at $E_{\nu}=1.5$~GeV, the contribution from $S_{11}(1535)$ resonance is about 95\%, which becomes about 90\% at $E_{\nu}=3$~GeV. Similar observation for the $S_{11}(1535)$ resonance dominance has been made in the case of neutrino induced $\eta$ production from the proton target. However, in the case of antineutrino induced reaction on the proton target, the contribution from $S_{11}(1535)$ resonance is about 88\% at $E_{\bar{\nu}}=1.5$~GeV, which becomes 84\% at $E_{\bar{\nu}}=3$~GeV, while in the case of antineutrino induced reaction off the neutron target, the contribution from $S_{11}(1535)$ resonance is about 94\% at $E_{\bar{\nu}}=1.5$~GeV, which becomes 85\% at $E_{\bar{\nu}}=3$~GeV.
 
 To understand the relative magnitude of the total cross section induced by the charged and neutral current reactions, in Fig.~\ref{Ch12_fg_eta:cc_xsec_ratio}, we have shown the results for the ratio of the total cross section for the charged current induced $\nu_{\mu}$ and $\bar{\nu}_{\mu}$ scattering on neutron and proton targets, respectively, to the cross section of the corresponding neutral current reactions on the isoscalar nucleon target, i.e., $\frac{\sigma_{\nu(\bar{\nu})p} + \sigma_{\nu(\bar{\nu})n}}{2}$. It may be noticed that this ratio increases with energy until $E_{\nu}=2$~GeV, after which it saturates to 8.25~(8.5) for neutrino~(antineutrino) reactions. Therefore, a constant factor for $\sigma(CC):\sigma(NC)$ ratio should not be considered for the (anti)neutrino experiments, where the average energy lies in the sub-GeV  region like at MicroBooNE, T2K, etc.
 
 In Fig.~\ref{fg_eta:cc_q2_weak}, we have presented the results for the $Q^2$ distribution~(i.e. $\frac{d\sigma}{dQ^2}$) vs $Q^2$ and $p_{\eta}$ distribution~(i.e. $\frac{d\sigma}{dp_{\eta}}$) vs. $p_{\eta}$ using the full model, for the charged current $\nu_{\mu}$ induced $\eta$ production from the free neutron target 
 at $E_{\nu_{\mu} } =1$, 1.5 and 4 GeV. Notice that different scale factors for $Q^2$ and $p_{\eta}$ distributions have been used to depict the results at $E_{\nu_{\mu}}=1$~GeV. 

\subsection{Flux averaged cross section}
\begin{figure*}
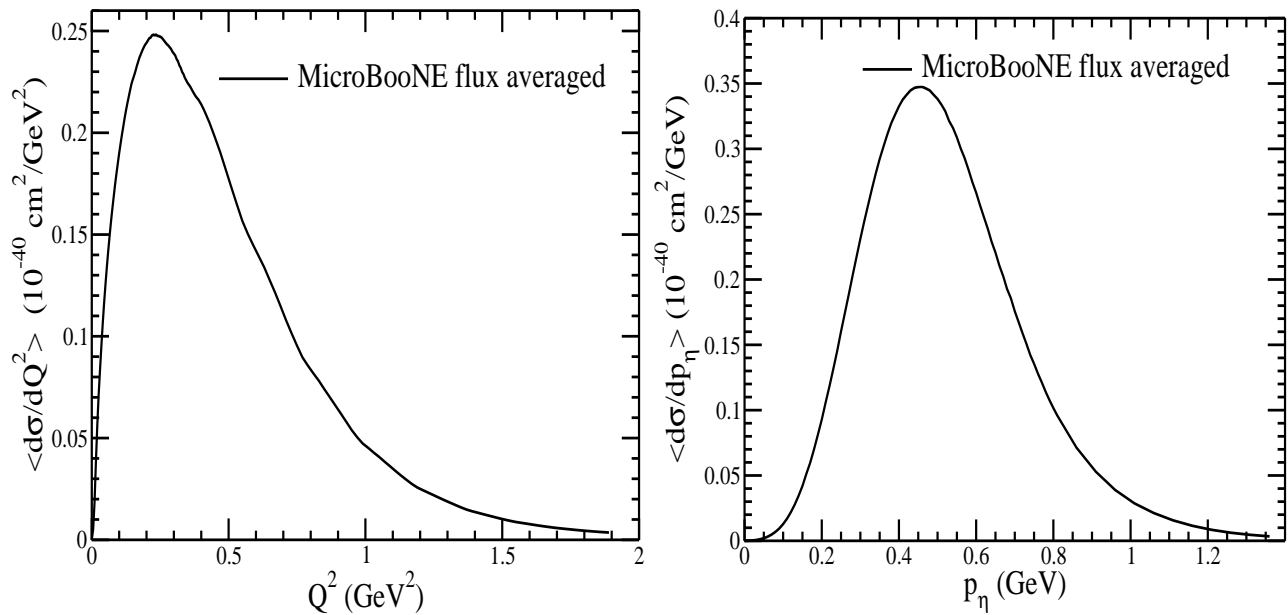

\begin{center}
\includegraphics[width=0.47\textwidth,height=.45\textwidth]{dsigma_dq2_neutrino_flux_average.eps}
\includegraphics[width=0.47\textwidth,height=.45\textwidth]{dsigma_dpeta_neutrino_flux_average.eps}
\caption{$\left\langle \frac{d\sigma}{dQ^2} \right \rangle$~(left panel) and $\left\langle \frac{d\sigma}{dp_{\eta}} \right\rangle$~(right panel) averaged over the MicroBooNE flux defined in Eq.~(\ref{dsig:flux}) for the charged current induced $\nu_{\mu} + n \longrightarrow \mu^{-} + p + \eta$ process.  }
\label{fg_eta:cc_q2_flux}
\end{center}
\end{figure*}
To explicitly see the $Q^2$ and $\eta$-momentum distribution at MicroBooNE energies, we have obtained the flux averaged differential and total scattering cross sections by folding it over the MicroBooNE flux~\cite{MicroBooNE:2019nio}. For this we define
\begin{eqnarray}\label{dsig:flux}
 \left \langle \frac{d\sigma}{dQ^2} \right \rangle &=& \frac{\int \frac{d\sigma}{dQ^2} \Phi (E_{\nu}) dE_{\nu} }{\int \Phi (E_{\nu}) dE_{\nu}}, \qquad \quad \left \langle \frac{d\sigma}{dp_{\eta}} \right \rangle = \frac{\int \frac{d\sigma}{dp_{\eta}} \Phi (E_{\nu}) dE_{\nu} }{\int \Phi (E_{\nu}) dE_{\nu}},
 \end{eqnarray}
 and  
 \begin{eqnarray}\label{sigma:flux}
  \langle \sigma \rangle &=& \frac{\int \sigma(E_{\nu}) \Phi (E_{\nu}) dE_{\nu} }{\int \Phi (E_{\nu}) dE_{\nu}}
\end{eqnarray}
where $\Phi(E_{\nu})$ is the MicroBooNE $\nu_{\mu}$ flux~\cite{MicroBooNE:2019nio}.

The results obtained for the flux averaged $Q^2$ and $\eta$-momentum distributions~(using Eq.~(\ref{dsig:flux})) for the charged current induced $\eta$ production by $\nu_{\mu}$ are shown in Fig.~\ref{fg_eta:cc_q2_flux}.
Using Eq.~(\ref{sigma:flux}), we obtain the charged current $\nu_{\mu}$ induced total cross section averaged over the MicroBooNE flux  to be $\langle\sigma_{CC}\rangle = 1.68 \times 10^{-41}$ cm$^{2}$.  We have also obtained the flux averaged cross section for the neutral current induced reactions $\nu p \rightarrow \nu p \eta$ and $\nu n \rightarrow \nu n \eta$, for which the results are found to be   $\langle\sigma_{NC(\nu p)}\rangle = 0.18 \times 10^{-41}$ cm$^{2}$ and $\langle\sigma_{NC(\nu n)}\rangle = 0.26 \times 10^{-41}$ cm$^{2}$, respectively, which corresponds to an average NC cross section for an isoscalar nucleon target to be $\langle \sigma_{NC(\nu N)} \rangle = 0.22 \times 10^{-41}$ cm$^{2}$. 
As discussed earlier, the main source of uncertainty in the theoretical prediction of (anti)neutrino cross section off the nucleon target is due to the uncertainty in the axial-vector form factor. This arises due to the large uncertainty in the branching fraction of the resonance to $N\pi$ decay mode,  and the choice of axial dipole mass $M_{A}$. For example, a 15\% variation from the central value in the strong coupling strength $g_{RN\pi}$ for $S_{11}(1535)$ resonance  leads to an uncertainty of about 4\% in the total flux averaged cross section, and a 10\% variation in $M_{A}$ leads to a variation of about 5\% in the flux averaged cross section, which leads to a total uncertainty of $ 0.11 \times 10^{-41}$ cm$^{2}$~($ 0.014 \times 10^{-41}$ cm$^{2}$) in the  flux averaged cross section for the charged~(neutral) current induced $\eta$ production from the free nucleon target. 

The MicroBooNE collaboration has reported the results for $\langle \sigma \rangle = 3.22 \pm 0.84 \pm 0.86 \times 10^{-41}$ cm$^2$/nucleon in argon nuclear target, where nuclear medium and final state interaction of $\eta$ mesons with the residual nucleus effects are also important. This needs to be taken into account which has been shown to be important in the case of photo- and electro- production of $\eta$ mesons~\cite{Lehr:2003ka, Lehr:2003km, Lee:1996gu}. This work is in progress and would be reported in future communication. 


\section{Summary and conclusions}\label{conclude}
We have studied the charged and neutral current $\nu_l({\bar\nu}_l);~(l=e,\mu)$  induced $\eta$ production off the nucleons and presented the results for the total scattering cross section $\sigma(E_{\nu_{l}(\bar{\nu}_{l})})$, $Q^2$-distribution~$\left(\frac{d\sigma}{dQ^2}\right)$ and the momentum distribution~$\left(\frac{d\sigma}{dp_\eta}\right)$ for the $\eta$ mesons, in a model in which the contribution from the nonresonant Born terms and the resonant terms are calculated in an effective Lagrangian approach. 
 We have applied this model to obtain the flux averaged differential and total scattering cross sections for the MicroBooNE $\nu_{\mu}$ flux.
 \vspace{2mm}
 
 We find that:
 
 \begin{itemize}
  \item [(i)] Weak charged current production of $\eta$ mesons induced by $\nu_{l}$ and $\bar{\nu}_{l}$~($l=e,\mu$) from the free nucleon target is dominated by the excitation of $S_{11}(1535)$ resonance and its subsequent decay into $\eta$ through $S_{11}(1535) \rightarrow N\eta$ decay, similar to the observations made in the case of electromagnetic production of $\eta$ mesons.
  
  \item [(ii)] This dominance of $S_{11}(1535)$ resonance contribution in the weak production of $\eta$ occurs in the charged as well as the neutral current induced reactions. However, at higher neutrino energies~($E_{\nu}> 2$~GeV), the contribution from the higher resonances becomes non-negligible.
  
  \item [(iii)] The charged as well as neutral current productions of $\eta$ meson are dominated by the vector current contribution.
  
  \item [(iv)] Weak charged current production cross section of $\eta$ meson is larger for the neutron target than the proton target. This is expected because $\eta$ production from neutron is induced by neutrinos while on the proton target, it is induced by the antineutrinos.
  
  \item [(v)] In the case of neutral current induced $\eta$ production, the cross section is larger from the neutron as compared to the proton target. 
  This is due to the isospin structure of the neutral current in the standard model.
  
  \item [(vi)] The charged current production cross section of the $\eta$ meson is larger than the neutral current production cross section. The enhancement factor is neutrino energy dependent. 
  For example, this ratio is 4:1 at $E_{\nu_{\mu}}=1$~GeV and becomes 8:1 at $E_{\nu_{\mu}}=2$~GeV. Similar observation has also been made in the case of antineutrino induced reactions.  
   
  \item [(vii)] The total scattering cross section folded over the MicroBooNE $\nu_{\mu}$ flux  is obtained to be $1.68\times 10^{-41}$ cm$^{2}$ and $0.22 \times 10^{-41}$ cm$^{2}$, respectively, for the charged and neutral current induced $\eta$ production from the free nucleon.

 \end{itemize}
 
 To conclude, the results presented, in this work, for the neutral and charged current induced (anti)neutrino scattering cross section from the free nucleon, ratio of the cross sections for the charged current to neutral current, and the flux averaged total cross section $\langle \sigma \rangle$, differential cross sections $\langle \frac{d\sigma}{dQ^2}\rangle$ and $\langle \frac{d\sigma}{dp_\eta} \rangle$ integrated over the MicroBooNE $\nu_{\mu}$ spectrum may be useful in the future analysis of MicroBooNE as well as other accelerator and atmospheric neutrino experiments like T2K, NOvA, DUNE, HyperK, etc. being performed in the few GeV energy region.

\section*{Acknowledgements}
We are thankful to D. Caratelli for many useful discussions regarding the $\eta$ production analysis being done at the MicroBooNE experiment. 
 AF and MSA are thankful to the
Department of Science and Technology (DST), Government of India for providing financial assistance under Grant No.
SR/MF/PS-01/2016-AMU.

 \bibliographystyle{apsrev}

\begin{thebibliography}{65}
\bibitem{SajjadAthar:2022pjt}
M.~Sajjad Athar, A.~Fatima and S.~K.~Singh, 
Prog. Part. Nucl. Phys. \textbf{129}, 104019 (2023)
doi:10.1016/j.ppnp.2022.104019

\bibitem{Athar:2020kqn}
M.~Sajjad Athar and S.~K.~Singh,
Cambridge University Press, 2020,
ISBN 978-1-108-77383-6, 978-1-108-48906-5
doi:10.1017/9781108489065

\bibitem{MicroBooNE}
    https://microboone.fnal.gov/
    
    

\bibitem{Machado:2019oxb}
P.~A.~Machado, O.~Palamara and D.~W.~Schmitz,
Ann. Rev. Nucl. Part. Sci. \textbf{69}, 363-387 (2019)
doi:10.1146/annurev-nucl-101917-020949.

\bibitem{T2K}
    https://t2k-experiment.org/
    
    

\bibitem{Hyper-Kamiokande:2022smq}
J.~Bian \textit{et al.} [Hyper-Kamiokande],
[arXiv:2203.02029 [hep-ex]].

\bibitem{DUNE:2022aul}
A.~Abed Abud \textit{et al.} [DUNE],
[arXiv:2203.06100 [hep-ex]].

\bibitem{Hyper-Kamiokande:2018ofw}
K.~Abe \textit{et al.} [Hyper-Kamiokande],
[arXiv:1805.04163 [physics.ins-det]].

\bibitem{JUNO}
http://juno.ihep.cas.cn/

\bibitem{INO}
https://www.ino.tifr.res.in/ino/

\bibitem{Fatima:2020tyh}
A.~Fatima, Z.~Ahmad Dar, M.~Sajjad Athar and S.~K.~Singh,
Int. J. Mod. Phys. E \textbf{29}, 2050051 (2020)
doi:10.1142/S0218301320500512
 

\bibitem{RafiAlam:2010kf} 
  M.~Rafi Alam, I.~Ruiz Simo, M.~Sajjad Athar and M.~J.~Vicente Vacas,
    Phys.\ Rev.\ D {\bf 82}, 033001 (2010)
  doi:10.1103/PhysRevD.82.033001
  

\bibitem{RafiAlam:2012bro}
M.~Rafi Alam, I.~Ruiz Simo, M.~Sajjad Athar and M.~J.~Vicente Vacas,
Phys. Rev. D \textbf{87}, 053008 (2013)
doi:10.1103/PhysRevD.87.053008

\bibitem{Dombey:1968vh} 
  N.~Dombey,
    Phys.\ Rev.\  {\bf 174}, 2127 (1968) 
  doi:10.1103/PhysRev.174.2127
  

\bibitem{Alam:2013vwa} 
  M.~Rafi Alam, M.~Sajjad Athar, L.~Alvarez-Ruso, I.~Ruiz Simo, M.~J.~Vicente Vacas and S.~K.~Singh,
    arXiv:1311.2293 [hep-ph]
  

\bibitem{Nakamura:2015rta}
S.~X.~Nakamura, H.~Kamano and T.~Sato,
Phys. Rev. D \textbf{92}, 074024 (2015)
doi:10.1103/PhysRevD.92.074024


\bibitem{Fatima:2022nfn}
A.~Fatima, M.~Sajjad Athar and S.~K.~Singh,
Phys. Rev. D \textbf{107},  033002 (2023)
doi:10.1103/PhysRevD.107.033002.

\bibitem{BEBCWA59:1989ofp}
W.~Wittek \textit{et al.} [BEBC WA59],
Z. Phys. C \textbf{44}, 175 (1989)
doi:10.1007/BF01557323

\bibitem{ICARUS}
I. Kochanek, PhD, Silesia University, Katowice (2015).

\bibitem{MicroBooNE:2023dqf}
P.~Abratenko \textit{et al.} [MicroBooNE],
[arXiv:2305.16249 [hep-ex]].

\bibitem{Weinberg:1967tq}
S.~Weinberg,
Phys. Rev. Lett. \textbf{19}, 1264 (1967)
doi:10.1103/PhysRevLett.19.1264

\bibitem{Salam:1968rm}
A.~Salam,
Conf. Proc. C \textbf{680519}, 367 (1968)
doi:10.1142/9789812795915\_0034

\bibitem{CrystalBallatMAMI:2010slt}
E.~F.~McNicoll \textit{et al.} [Crystal Ball at MAMI Collaboration],
Phys. Rev. C \textbf{82}, 035208 (2010)
[erratum: Phys. Rev. C \textbf{84}, 029901 (2011)]
doi:10.1103/PhysRevC.84.029901

\bibitem{A2:2014pie}
D.~Werthm\"uller \textit{et al.} [A2 Collaboration],
Phys. Rev. C \textbf{90},  015205 (2014)
doi:10.1103/PhysRevC.90.015205

\bibitem{Denizli:2007tq} 
  H.~Denizli {\it et al.} [CLAS Collaboration],
    Phys.\ Rev.\ C {\bf 76}, 015204 (2007)
  doi:10.1103/PhysRevC.76.015204
 

\bibitem{DiDonato:2011kr}
C.~Di Donato, G.~Ricciardi and I.~Bigi,
Phys. Rev. D \textbf{85}, 013016 (2012)
doi:10.1103/PhysRevD.85.013016

\bibitem{Faessler:2008ix}
A.~Faessler, T.~Gutsche, B.~R.~Holstein, M.~A.~Ivanov, J.~G.~Korner and V.~E.~Lyubovitskij,
Phys. Rev. D \textbf{78}, 094005 (2008)
doi:10.1103/PhysRevD.78.094005


\bibitem{ParticleDataGroup:2020ssz}
P.~A.~Zyla \textit{et al.} [Particle Data Group],
PTEP \textbf{2020},  083C01 (2020)
doi:10.1093/ptep/ptaa104

\bibitem{Donnachie:1978fm}
A.~Donnachie and G.~Shaw,
``Electromagnetic Interactions of Hadrons. 1.,''
Springer, New York 1978, 446p.

\bibitem{Amaldi:1979vh} 
  E.~Amaldi, S.~Fubini and G.~Furlan,
    Springer Tracts Mod.\ Phys.\  {\bf 83}, 1 (1979)
  doi:10.1007/BFb0048209
  

\bibitem{Drechsel:1994zx}
D.~Drechsel,
Few Body Syst. Suppl. \textbf{7}, 325 (1994)

\bibitem{Bradford:2006yz}
  R.~Bradford, A.~Bodek, H.~S.~Budd and J.~Arrington,
    Nucl.\ Phys.\ Proc.\ Suppl.\  {\bf 159} (2006) 127
  doi:10.1016/j.nuclphysbps.2006.08.028
  


\bibitem{Tiator:2011pw} 
  L.~Tiator, D.~Drechsel, S.~S.~Kamalov and M.~Vanderhaeghen,
    Eur.\ Phys.\ J.\ ST {\bf 198}, 141 (2011)
  doi:10.1140/epjst/e2011-01488-9
 


\bibitem{Bernard:2001rs} 
  V.~Bernard, L.~Elouadrhiri, U.~G.~Meissner,
    J.\ Phys.\ G {\bf 28}, R1 (2002)
    doi:10.1088/0954-3899/28/1/201
    
    

\bibitem{MicroBooNE:2019nio}
P.~Abratenko \textit{et al.} [MicroBooNE],
Phys. Rev. Lett. \textbf{123},  131801 (2019)
doi:10.1103/PhysRevLett.123.131801

\bibitem{Lehr:2003ka}
J.~Lehr and U.~Mosel,
Phys. Rev. C \textbf{68}, 044603 (2003)
doi:10.1103/PhysRevC.68.044603

\bibitem{Lehr:2003km}
J.~Lehr, M.~Post and U.~Mosel,
Phys. Rev. C \textbf{68}, 044601 (2003)
doi:10.1103/PhysRevC.68.044601

\bibitem{Lee:1996gu}
F.~X.~Lee, L.~E.~Wright, C.~Bennhold and L.~Tiator,
Nucl. Phys. A \textbf{603}, 345-366 (1996)
doi:10.1016/0375-9474(96)80006-D


\end{thebibliography}

\end{document}